\documentclass[12pt]{iopart}

\expandafter\let\csname equation*\endcsname\relax

\expandafter\let\csname endequation*\endcsname\relax
\usepackage{amsmath}
\usepackage{amsfonts}
\usepackage{braket}
\usepackage{cite}
\usepackage{amssymb}

\usepackage{graphicx}
\usepackage[font=small,labelfont=bf,
   justification=justified,singlelinecheck=false, format=plain]{caption}

\usepackage[colorlinks, citecolor = blue, urlcolor = blue]{hyperref}
\begin{document}

\title[Reaction-diffusion dynamics of the weakly dissipative Fermi gas]{Reaction-diffusion dynamics of the weakly dissipative Fermi gas}

\author{Hannah Lehr$^1$, Igor Lesanovsky$^{1,2}$, Gabriele Perfetto$^{1,*}$}
\address{$^1$ Institut f\"{u}r Theoretische Physik,  Universit\"{a}t T\"{u}bingen, Auf der Morgenstelle 14, 72076 T\"{u}bingen, Germany}
\address{$^2$ School of Physics and Astronomy and Centre for the Mathematics and Theoretical Physics of Quantum Non-Equilibrium Systems, The University of Nottingham, Nottingham, NG7 2RD, United Kingdom}
\address{$*$ Author to whom any correspondence should be addressed.}
\ead{gabriele.perfetto@uni-tuebingen.de}

\begin{abstract}
We study the one-dimensional Fermi gas subject to dissipative reactions. The dynamics is governed by the quantum master equation, where the Hamiltonian describes coherent motion of the particles, while dissipation accounts for irreversible reactions. For lattice one-dimensional fermionic systems, emergent critical behavior has been found in the dynamics in the reaction-limited regime of weak dissipation. Here, we address the question whether such features are present also in a gas in continuum space. We do this in the weakly dissipative regime by applying the time-dependent generalized Gibbs ensemble method. We show that for two body $2A\to \emptyset$ and three $3A\to \emptyset$ body annihilation, as well as for coagulation $A+A\to A$, the density features an asymptotic algebraic decay in time akin to the lattice problem. In all the cases, we find that upon increasing the temperature of the initial state the density decay accelerates, but the asymptotic algebraic decay exponents are not affected. We eventually consider the competition between branching $A\to A+A$ and the decay processes $A\to \emptyset$ and $2A\to \emptyset$. We find a second-order absorbing-state phase transition in the mean-field directed percolation universality class. This analysis shows that emergent behavior observed in lattice quantum reaction-diffusion systems is present also in continuum space, where it may be probed using ultra-cold atomic physics.

\end{abstract}


\section{Introduction}
Nonequilibrium phenomena, where time-reversal invariance is broken and transport of particles over macroscopic scales takes place, display a variety of behaviors which is hard to encompass into universal principles \cite{kogut,wilson,cardyscaling,huang2008statistical}. Within this timely domain of research, reaction-diffusion (RD) models \cite{vladimir1997nonequilibrium, hinrichsen2000non, henkel2008non, tauber2002dynamic, tauber2005, tauber2014critical} stand out as paradigmatic nonequilibrium systems where universal principles characterizing their dynamics can be uncovered. 

In classical RD models, particles are diffusively transported in space (diffusion constant $D$) in a discrete or continuous space of dimension $d$. Upon meeting, particles undergo reaction processes at rate $\Gamma$, that either destroy or produce new particles. Typical examples of reactions depleting the system are binary $2A\to \emptyset$ and three-body $3A\to\emptyset$ annihilation, or coagulation $2A\to A$ of two particles merging into a single one. In the case where only such reactions are present, the system evolves towards the vacuum state devoid of particles. In the ``diffusion-limited'' regime $\Gamma/D \gg 1$, the approach towards the vacuum state at large time scales displays universal algebraic decay $n\sim (D t)^{\delta}$ with critical exponent $\delta<0$. Universality here means that the asymptotic decay is independent of the microscopic details, while it, instead, depends only on macroscopic quantities, such as the diffusion constant $D$, and on the space dimensionality $d$. The diffusion-limited decay is independent from the reaction rate $\Gamma$ since nearby particles in this limit rapidly react. The asymptotic decay is then controlled by the appearance of particle anticorrelations due to large ``depletion zones'' separating far apart particles. The time needed to cross such empty areas is dictated by the diffusion constant $D$ only, hence the density decay takes a universal scaling form. For binary annihilation and coagulation, for example, one finds $\delta=-1/2$ in $d=1$ \cite{toussaint1983particle, Racz1985, spounge88, torney1983diffusion, privman94,Peliti_1986, Lee1994,cardy1996, tauber2002dynamic, tauber2005, tauber2014critical}. This kinetics deviates from the mean-field prediction $n(t)\sim (\Gamma t)^{-1}$. The latter is recovered only in $d>2$ or in the so-called ``reaction-limited'' regime $\Gamma/D \ll 1$, when diffusion is effective in filling the whole available space \cite{kang84scaling,kang85,kang84fluct,vladimir1997nonequilibrium,henkel2008non}. Note that in the reaction-limited regime, the asymptotic decay is dictated by the reaction rate $\Gamma$ only, while it is independent from $D$, contrary to the diffusion-limited case. This result comes from the fast diffusive mixing, which wipes out spatial fluctuations, such as depletion zones. In the driven case, also  processes that create particles are present. A typical example is the so-called contact process, where branching of a particle $A\to A+A$ (rate $\Gamma_{\beta}$) into a nearby offspring competes with single-body decay $A\to\emptyset$ (rate $\Gamma_{\delta}$). When the relative strength $\Gamma_{\beta}/\Gamma_{\delta}>(\Gamma_{\beta}/\Gamma_{\delta})_c$ between the two processes is poised above a critical value $(\Gamma_{\beta}/\Gamma_{\delta})_c$, one has an active stationary state, i.e., a stationary state with non-zero particle density $n^{\mathrm{stat}}\neq 0$. Viceversa for $\Gamma_{\beta}/\Gamma_{\delta}<(\Gamma_{\beta}/\Gamma_{\delta})_c$ a vacuum-absorbing state with $n^{\mathrm{stat}}=0$ is found. The latter is a fluctuationless state which once dynamically accessed it can no longer be left. As such it manifestly violates detailed balance, thereby highlighting the nonequilibrium character of the associated stationary state transition. The latter for the contact process is a second-order phase transition, with algebraic behavior of the order parameter $n^{\mathrm{stat}}$ as a function of $\Gamma_{\beta}/\Gamma_{\delta}$ and time [at the critical point $(\Gamma_{\beta}/\Gamma_{\delta})_c$]. The associated critical exponents are found in the directed percolation universality class \cite{hinrichsen2000non,vladimir1997nonequilibrium,henkel2008non,tauber2014critical,Malte_Henkel_2004}. 

The RD dynamics has been generalized to quantum many-body systems. A motivation to do this is that quantum reaction-diffusion systems naturally model cold atoms subject to dissipative atomic losses and creations \cite{lossexp0,lossexp1,lossexp2,lossexp3,lossexp4,lossexp5,lossexp6,lossexp7,lossexp8,lossexpF1,lossexpF2,lossexpF3} due to the coupling with the surrounding environment. In addition, quantum RD systems naturally connect to kinetically constrained models \cite{diehl2008quantum, diehl2011topology, tomadin2011, carollo2022quantum, lesanovsky2013, olmos2014, marcuzzi16, gillman20}. In quantum RD, classical diffusion is replaced with coherent quantum hopping (rate $\Omega$). The latter gives ballistic transport. We adopt anyhow the name quantum RD henceforth, since this allows to link these quantum models to the classical ones. As a matter of fact, in spite of the different transport mechanism, quantum RD models also show asymptotic algebraic decay in time of the particle density and the same mean-field dynamics as classical RD. The quantum RD dynamics is then described by the quantum master equation of Lindblad form \cite{gorini1976completely, lindblad1976generators}, where reactions are implemented by quantum jump operators. The ensuing open-quantum dynamics is, however, not exactly solvable and therefore investigations in this field has been primarily conducted at the numerical level for small system sizes \cite{van_horssen2015,carollo2019, gillman2019, jo2021, carollo2022nonequilibrium}. 

Only recently, a full analytical treatment of quantum RD systems has been carried on in the thermodynamic limit in the reaction-limited regime of weak dissipation $\Gamma/\Omega \ll 1$. This analysis is based on the time-dependent generalized Gibbs ensemble (TGGE) method \cite{TGGE1,TGGE2,TGGE3,TGGE4,TGGE5,TGGE6}. This approach relies on the separation of time scales between unitary evolution and dissipation. The fast Hamiltonian dynamics leads to fast relaxation to a state that is constructed, according to the maximal entropy principle, in the generalized Gibbs form \cite{GGE1,GGE2}, which takes into account the extensive number of conservation laws of the Hamiltonian. From the latter state predictions on the long-time behavior of the particle density can be obtained by deriving a kinetic equation, valid for perturbatively weak reactions, for the occupation function in momentum space. This approach has been specifically pursued in a variety of lattice models ranging from spin systems \cite{riggio2023,ali2024signatures,TGGEnumMario}, bosons \cite{lossth3,rowlands2023quantum} and fermions \cite{lossth1,lossth6,perfetto22,perfetto23,lossth9}. For fermionic lattice systems, it has been shown that the quantum reaction-limited regime displays critical nonequilibrium behavior much richer than in the classical RD case. This manifests, for example, into algebraic decay $n(t) \sim (\Gamma t)^{-1/2}$ for $2A\to\emptyset$ with exponent different from mean field \cite{lossth1,perfetto22} and from that $n(t)\sim (\Gamma t)^{-1}$ of coagulation $2A\to A$, unlike in the classical case. Quantum coherent effects can even more dramatically impact the long-time dynamics by rendering the particle decay not algebraic, as in the case of $3A\to \emptyset$ \cite{perfetto23}, or the stationary state associated to absorbing-state phase transition quantum correlated as a consequence of locally-protected dark states \cite{perfetto22}. Recently, in Ref.~\cite{TGGEnumMario}, the TGGE method has been also successfully compared with numerical tensor-network simulations of the dynamics of spin $1/2$ systems with weak single-particle decay and creation.   

The case of fermionic gases in continuum space has been, so far, discussed in Refs.~\cite{Rosso2022,gerbino2023largescale,gerbino2024kinetics,maki2024loss} only for binary annihilation $2A\to\emptyset$. Therein the density shows the same algebraic decay $n(t)\sim (\Gamma t)^{-1/2}$ in $d=1$ as in the aforementioned lattice case. The goal of this manuscript is precisely to address whether for more complicated processes, such as higher-body decay $3A\to \emptyset$, coagulation $2A\to A$ and particle-creating processes $A\to 2A$, emergent behavior in quantum reaction-diffusion models is impacted on by the continuum-space limit. We show that, in quantum RD, the exponents associated to the relaxation algebraic decay and to the stationary state phase transition are robust with respect to the formulation in terms of a gas in continuum space. We do this by studying the Fermi gas in one spatial dimension in the thermodynamic limit and in the reaction-limited regime of weak dissipation. First, we show how the quantum master equation in continuum space can be obtained by taking the limit of vanishing lattice spacing in the associated lattice formulation.
Once the continuum limit of the jump operators embodying the reaction processes is established, we apply the TGGE method to study the dynamics. 

We start by benchmarking the continuum gas formulation by studying binary annihilation $2A\to \emptyset$. Therein the continuum limit of the jump operators is simple to establish as the latter is formed only by commuting annihilation operators. From this approach, we therefore retrieve the dissipative kinetic equation of Refs.~\cite{Rosso2022,gerbino2023largescale,gerbino2024kinetics,maki2024loss} and thereby a decay $n(t)\sim (\Gamma t)^{-1/2}$ for zero-temperature Fermi-sea initial states. Importantly, we also extend previous studies by accounting for the experimentally relevant case of initial state having a finite temperature $T$. We find that the latter affects the amplitude of the decay but not its exponent. In particular, as the temperature is increased, we observe that the decay becomes increasingly faster according to the law $n(t)\sim (\Gamma T t)^{-1/2}$, i.e., with the temperature rescaling time. This is starkly different from the lattice counterpart of the problem. In the latter the density decays with mean-field exponent $n(t)\sim (\Gamma t)^{-1}$ in an initial transient time window whose width increases with $T$, and thereby approaches mean field for large temperature values. This different behavior is  physically explained by the fact that for a gas in continuum space high temperatures can excite very high velocities among the particles. Faster moving particles mix more rapidly leading to an accelerated decay. On the lattice, on the other hand, particles have a maximal propagation velocity. Large temperatures then simply lead to a maximal entropy homogeneous reshuffling of the particles among the allowed modes. This destroys correlations and leads to mean-field decay in the reaction-limited regime. We also consider the case of three-body annihilation $3A\to \emptyset$, for which we compute the effective decay exponent and we find that it does not converge in time meaning that the ensuing decay is not algebraic. This result shows that the corrections to the algebraic decay found in Ref.~\cite{perfetto23} are not a lattice effect and persist in the continuum gas. Moving to the case of coagulation decay $2A\to A$, one finds that the continuum limit of the latter jump operator requires additional care compared to $K-$body losses since it contains non-commuting creation and annihilation operators. The continuum limit is then taken by first normal ordering the jump operator and then by introducing an ultraviolet (UV), short distance, cutoff. From the resulting kinetic equation, we find mean-field algebraic decay $n(t) \sim t^{-1}$, with an amplitude which depends on the UV cutoff. Also for coagulation, we consider the case of finite temperature $T$ initial states and we do find that only the amplitude of the decay is impacted on by $T$, but not the exponent. In particular, also in continuum space binary annihilation and coagulation display algebraic decays with different exponents in contrast with the classical case. Finally we consider a contact process model obtained via the competition between branching $A\to 2A$, decay $A\to \emptyset$ and binary annihilation $2A\to \emptyset$. In classical RD systems \cite{vladimir1997nonequilibrium,hinrichsen2000non,henkel2008non, tauber2002dynamic, tauber2005, tauber2014critical}, binary annihilation is included only to avoid the particle density to diverge in the active state. For fermionic systems the double occupancy restriction necessarily makes the stationary density finite and therefore binary annihilation is expected to be superfluous. In Ref.~\cite{perfetto22}, however, it has been shown for the fermionic lattice counterpart of the model that local dark states of the binary annihilation reaction lead to the emergence of stationary correlations beyond mean field. Motivated by this result, we study in this work whether also in continuum space binary annihilation is necessary to induce stationary correlations. Also in the case of branching, the associated continuum limit requires the introduction of UV cutoff. The latter affects only the value of the critical point $(\Gamma_{\beta}/\Gamma_{\delta})_c$, but not the critical exponents associated to transition. This further confirms that the value of the critical-point $(\Gamma_{\beta}/\Gamma_{\delta})_c$ is a non-universal feature, while the long-time behavior is universal. The transition that we find is, indeed, in the mean-field directed percolation universality class, as in the lattice case. A crucial difference from the latter is, however, present in the spatial correlations in the stationary state. We, indeed, find that correlations beyond mean field are generically present even in the absence of binary annihilation. The contribution of the lattice dark states of binary annihilation to stationary correlations found in Ref.~\cite{perfetto22} is therefore a lattice effect which disappears in the continuum limit. In the latter case, binary annihilation is therefore unimportant since the universality class of the stationary state phase transition and the structure of the associated stationary state are determined only by decay and branching, as well as in classical RD.

The manuscript is organized as follows. In Sec.~\ref{sec:II_continuum}, we introduce quantum RD systems via Lindblad quantum master equation. In Sec.~\ref{sec:III_GGE}, we discuss the TGGE method to tackle the weakly-dissipative reaction-limited regime. Section \ref{sec:IV_results} contains the results of the manuscript. We present therein the cases of binary annihilation $2A\to \emptyset$, three-body annihilation $3A\to\emptyset$, coagulation $2A\to A$ and branching $A\to 2A$. In Sec.~\ref{sec:V_conclusions}, we eventually draw our final conclusions. Details about the calculations are reported in the Appendices.

\section{Model and continuum limit}
\label{sec:II_continuum}
We define in this Section the fermionic quantum RD models. In Subsec.~\ref{subsec:lattice}, we formulate the quantum master equation for one-dimensional lattice systems, with lattice spacing $a$. In Subsec.~\ref{subsec:continuum_gas}, we take the limit of vanishing lattice spacing $a\to 0$ to obtain the continuum gas description.

\subsection{\textbf{Lattice model}}
\label{subsec:lattice}
The dynamics is modeled via a phenomenological quantum master equation, which we choose to be in the Markovian form ($\hbar=1$ henceforth): \cite{gorini1976completely, lindblad1976generators} 
\begin{equation}
\dot{\rho}(t)=-i[H,\rho(t)]+\mathcal{D}[\rho(t)].
\label{eq:master_equation}   
\end{equation}
Here, $H$ denotes the Hamiltonian, while $\mathcal{D}$ embodies dissipative processes due to the coupling with the surrounding environment. In the lattice case, $H$ is a quadratic tight-binding hopping Hamiltonian 
\begin{equation}
H=-\overline{\Omega}\sum_{j=1}^{N}(c_{a j}^{\dagger} c_{aj+a}+c_{aj+a}^{\dagger}c_a) \, .
\label{eq:free_fermion_Hamiltonian_lattice}
\end{equation}
with $\overline{\Omega}$ the hopping amplitude which has the units of inverse time. Here, $L=N a$ is the length of the chain, with $N$ the number of sites and $a$ the lattice spacing. We adopt periodic boundary conditions $c_{aj +L}=c_{a j}$ and we have the standard fermionic anticommutation relations $\{c_{a j}, c_{a j'}^{\dagger} \}=\delta_{j,j'}$. We explicitly report the lattice spacing $a$ length as it will be useful to take the continuum limit. Throughout, we consider only the case where both the Hamiltonian \eqref{eq:free_fermion_Hamiltonian_lattice} and the dissipator $\mathcal{D}$ in Eq.~\eqref{eq:master_equation}, and therefore the whole dynamics, are translational invariant.

The dissipator $\mathcal{D}$ on the lattice takes the Lindblad form 
\begin{equation}
\mathcal{D}[\rho]=\sum_{j,\nu} \left[L_{a j}^{\nu}\rho {L_{a j}^{\nu}}^\dagger-\frac{1}{2}\left\{{L_{a j}^{\nu}}^\dagger L_{a j}^{\nu},\rho \right\}\right].
\label{eq:dissipator}    
\end{equation}
The index $\nu$ distinguishes the different dissipative processes present in the dynamics. The first and simplest of such process is one-body decay, $\nu=\delta$ rate $\Gamma_{\delta}$ [units: $\mbox{time}^{-1}$], $A\to \emptyset$:
\begin{equation}
L_{a j}^{\delta} = \sqrt{\Gamma_{\delta}} c_{a j}.
\label{eq:death}
\end{equation}  
When only this jump operator is considered, the Lindbladian \eqref{eq:master_equation} is quadratic and exactly solvable. A more interesting and nonlinear jump operator is, instead, provided by binary annihilation, $\nu=2 \alpha$ rate $\Gamma_{2\alpha}$, of two neighboring particles  $2A\to \emptyset$:
\begin{equation}
L_{a j}^{2\alpha}=\sqrt{\Gamma_{2\alpha}}\,c_{a j} c_{aj+a}.
\label{eq:2annihilation}
\end{equation}
This process implements a two-body loss in cold atoms \cite{lossexpF1,lossexpF2,lossexpF3}. We will also consider a modification of the jump operator \eqref{eq:2annihilation}, which allows for superposition of two decay channels. This is parametrized via an angle $\theta$: 
\begin{equation}
L_{a j}^{2\alpha}(\theta)=\sqrt{\Gamma_{2\alpha}} c_{a j}(\cos\theta \, c_{aj+a}-\sin\theta \, c_{aj-a}),
\label{eq:annihilation_interference}
\end{equation}
so that for $\theta=0,\pi$ the binary annihilation \eqref{eq:2annihilation} is retrieved. Jump operator of this type was obtained in Refs.~\cite{lossexp0,lossth7}, where the Bose-Hubbard under strong two-body losses (Zeno regime) is studied. By applying a dissipative Schrieffer Wolff transformation at second order in the hopping-Hamiltonian rate, the Bose-Hubbard Lindbladian is mapped to that of fermions \eqref{eq:free_fermion_Hamiltonian_lattice} under dissipation of the form \eqref{eq:annihilation_interference} with $\theta=\pi/4$. The ensuing dynamics was studied in the weak dissipation limit $\Gamma_{2\alpha}\ll\Omega$ first in Ref.~\cite{lossth1} for $\theta=\pi/4$, and then in Ref.~\cite{perfetto22} for generic $\theta$ values including the limit $\theta=0,\pi$ of \eqref{eq:2annihilation}. Moving to higher-body reactions, one has three-body annihilation, $\nu=3\alpha$ and rate $\Gamma_{3\alpha}$, of three neighboring particles $3A\to\emptyset$: 
\begin{equation}
L_{a j}^{3\alpha}=\sqrt{\Gamma_{3\alpha}}\,c_{a j} c_{aj+a} c_{aj+2a}.
\label{eq:3annihilation}
\end{equation}
This process was experimentally realized for lattice cold-atomic models in Refs.~\cite{lossexp1,lossexp2,lossexp7}. Theoretically three-body annihilation of fermionic particles has been considered in Refs.~\cite{perfetto23} in the regime of weak dissipation $\Gamma_{3\alpha}\ll \Omega$. 

The next nonlinear decay process that will be considered is coagulation, $\nu=\gamma$ rate $\Gamma_{\gamma}$, of two nearby particles merging into a single one $2A\to A$:
\begin{equation}
L_{a j}^{\gamma\pm}=\sqrt{\Gamma_{\gamma}/2} \, : c_{a j} n_{aj\pm a} := \sqrt{\Gamma_{\gamma}/2} c^{\dagger}_{aj \pm a} c_{aj \pm a} c_{aj}. \label{eq:coagulation}
\end{equation}
Importantly, in the previous equation we have introduced the normal ordered product $:\dots:$ of a set of fermionic creation/annihilation operators. The normal ordered product is defined so that all creation operators lie on the left, while annihilation to the right. For fermions one must also multiply the operator by the sign of the permutation needed to produce the desired order. On the lattice normal ordering of the jump operators is not needed, but it will be required when taking the continuum limit $a\to 0$ of Eq.~\eqref{eq:coagulation}. In the case of binary \eqref{eq:2annihilation}-\eqref{eq:annihilation_interference} and three-body annihilation, normal ordering is never necessary, neither on the lattice nor in the continuum, since these jump operators contain only annihilation operators. The latter always anticommute (even when taken on the same lattice/space point). We also note that we are considering the case of symmetric coagulation, where particles can either coagulate to the left, $L_j^{\gamma +}$, or to the right $L_j^{\gamma-}$ with equal rates. All the processes modelled by Eqs.~\eqref{eq:death}-\eqref{eq:coagulation} deplete the system leading to a vacuum stationary state. In these cases, universal behavior is found in the decay at long times of the particle density $n_{aj}=\braket{c_{aj}^{\dagger} c_{aj}}$. In order to sustain a stationary state with a nonzero density of particles and for allowing the possibility of absorbing-state phase transitions \cite{hinrichsen2000non,vladimir1997nonequilibrium,henkel2008non,tauber2014critical,Malte_Henkel_2004}, we now introduce a branching process, $\nu=\beta$ in Eq.~\eqref{eq:dissipator} and rate $\Gamma_{\beta}$, where a particle produces a nearby offspring $A\to2A$: 
\begin{equation}
L_{a j}^{\beta\pm}=\sqrt{\Gamma_{\beta}/2}\, :c_{aj}^{\dagger}n_{aj\pm a}:= \sqrt{\Gamma_{\beta}/2}\, c_{a j}^{\dagger}c_{a j\pm a}^{\dagger} c_{a j \pm a}. 
\label{eq:branching}
\end{equation} 
Normal ordering is again required when taking the continuum limit. We consider symmetric branching, i.e., it takes place at the same rate in the right and left direction, as in the case of coagulation \eqref{eq:coagulation}.

\subsection{\textbf{Continuum model}}
\label{subsec:continuum_gas}
In taking the continuum limit, we send the lattice spacing $a \to 0$. Simultaneously we take the number of sites $N\to \infty$ so that the volume $L=N a$ stays finite. We replace discrete sum $\sum_{j}\to a^{-1} \int \mbox{d}x$ by the corresponding integral. Fields are rescaled according to 
\begin{equation}
    \frac{c_{a j}}{a^{1/2}} \to \psi(x), \quad
    \frac{c^\dagger_{a j}}{a^{1/2}} \to \psi^\dagger(x),
\label{eq:continuum_space_scaling}
\end{equation}
so that they obey the fermionic anticommutation relations $\{\psi(x), \psi^{\dagger}(y) \}=\delta(x-y)$. The continuum limit of the Hamiltonian yields:
\begin{equation}
    H = \int_{0}^L \mbox{d} x \psi^{\dagger}(x)(-\Omega \partial_x^2)\psi(x),\label{eq:free_hamiltonian_continuum}
\end{equation}
with $\Omega= a^2 \overline{\Omega}$. In continuum space $\Omega$ has therefore units of [$\mbox{length}^2 \,\, \mbox{time}^{-1}$]. These are the units of a diffusion constant. It is, however, important to stress that in quantum mechanics the Hamiltonian \eqref{eq:free_hamiltonian_continuum} describes coherent ballistic transport of particles. This is a fundamental difference between classical RD systems, where particles are transported in space via  diffusion, and quantum RD systems where transport is ballistic. Note that the space integral now runs over the volume $L$ of the system.

Next, the continuum limit of the dissipator $\mathcal{D}$ in Eq.~\eqref{eq:dissipator} is taken
\begin{equation} 
\label{eq:dissipator_continuum}
    \mathcal{D}[\rho] = \sum_{\nu }\int_{0}^L \mbox{d} x [ L^{\nu}(x) \rho {L^{\nu}}^\dagger(x) - \frac{1}{2} \{ {L^{\nu}}^\dagger(x) L^{\nu}(x) , \rho \} ].
\end{equation}
Jump operators in the continuum are obtained from the lattice ones via
\begin{equation}
L^{\nu}(x)= \lim_{a\to 0} \frac{L^{\nu}_{a j}}{a^{1/2}},
\label{eq:jumps_continuum}
\end{equation}
where we express the lattice jump operators $L_{aj}^{\nu}$ in terms of the continuous fields \eqref{eq:continuum_space_scaling} and we expand by taking the leading term as $a\to 0$. This procedure, in general, also redefines the units of the associated rate (similarly to the case of the hopping amplitude $\overline{\Omega}\to \Omega$).

For the single body decay \eqref{eq:death}, one simply has 
\begin{equation}
L^{\alpha}(\vec{x})=\sqrt{\Gamma_{\delta}} \psi(x),
\label{eq:one_body_decay_continuum}
\end{equation}
with the rate $\Gamma_{\delta}$ coinciding with that of the lattice problem. Moving to the case of binary annihilation, one readily realizes that albeit \eqref{eq:2annihilation} and \eqref{eq:annihilation_interference} are different on the lattice for $\theta \neq 0,\pi$, they yield the same continuum limit for any $\theta$ \footnote{the case $\theta=\pi (4n-1)/4$, with $n\in \mathbb{Z}$ requires a separate treatment, but it does not change the asymptotic decay of the density, see Subsec.~\ref{subsec:2bresults} below and \ref{app:Appendix1}.}:
\begin{equation}
L^{2\alpha}(x) = \sqrt{\widetilde{\Gamma}_{2\alpha}} \psi (x)\partial_{x} \psi (x), \quad \mbox{with} \quad \widetilde{\Gamma}_{2\alpha}=a^3 \Gamma_{2\alpha}(\cos \theta +\sin \theta)^2.
\label{eq:binary_ann_continuum_FB}    
\end{equation}
Here $\widetilde{\Gamma}_{2\alpha}$ has therefore units of [$\mbox{length}^3 \, \, \mbox{time}^{-1}$]. We remark that on the lattice \eqref{eq:2annihilation} and \eqref{eq:annihilation_interference} can lead to different asymptotic decays for the density \cite{lossth1,perfetto22,riggio2023}, and correlations in the stationary state obtained when also branching \cite{perfetto22} is simultaneously considered. This is caused by the fact that the jump operator \eqref{eq:annihilation_interference} allowing for interference between two different decay channels acting on a classical product states creates superposition and coherences, while classical binary annihilation \eqref{eq:2annihilation} maps product states into product states and cannot create coherences on its own. In the continuum limit, however, the two processes are identical. The continuum limit of three-body annihilation \eqref{eq:3annihilation} can be similarly derived and it reads
\begin{equation}
L^{3\alpha}(x) = \sqrt{\widetilde{\Gamma}_{3\alpha}} \psi(x)\partial_x \psi(x) \partial^2_x \psi(x), \quad \mbox{with} \quad \widetilde{\Gamma}_{3\alpha}=4 a^8 \Gamma_{3\alpha}.
\label{eq:3ann_continuum}
\end{equation}
This procedure of taking the continuum limit can be extended to the generic case of $K-$body losses, which leads to the appearance of derivatives up to order $K-1$.

Finally, we consider the continuum limit of coagulation \eqref{eq:coagulation} and branching \eqref{eq:branching}, where normal ordering has to be taken into account. For coagulation we find
\begin{align}
L^{\gamma \pm}(x) &= \lim_{a \to 0}\frac{1}{\sqrt{a}}L_{a j}^\gamma  = \lim_{a\to 0} \sqrt{\frac{\Gamma_\gamma}{a}}c_{aj \pm a}^\dagger c_{aj \pm a} c_{aj} \nonumber \\
    &=\lim_{a \to 0} \sqrt{a^2\Gamma_\gamma}\psi^\dagger(x\pm a) \psi(x\pm a) \psi(x) \nonumber \\
    &=\lim_{a\to 0} \sqrt{a^2\Gamma_\gamma}\left(\psi^\dagger(x) \pm a\partial_x \psi^\dagger(x)+\mathcal{O}(a^2) \right) \left(\psi(x) \pm a\partial_x \psi(x) +\mathcal{O}(a^2)\right) \psi(x)  \nonumber \\
    &= \pm \sqrt{\widetilde{\Gamma}_{\gamma}}\psi^\dagger(x)\partial_x\psi(x) \psi(x) +\mathcal{O}(a^{3}), \quad \mbox{with} \quad  \widetilde{\Gamma}_{\gamma}= a^4\Gamma_\gamma.
\label{eq:coagulation_continuum}
\end{align}
Right and left coagulation therefore yield the same continuum limit since the overall sign of the jump operators is irrelevant. It is important to emphasize the importance of starting from the normal-ordered operator in Eq.~\eqref{eq:coagulation}. Normal ordering, indeed, allows to keep track of the correct ordering of the operators in $L_j^{\gamma}$ so that the creation operator is always applied after the annihilation ones. This is important as $a\to 0$ when the creation and annihilation operators in $L_j^{\gamma}$ get applied on the same point. If $L_j^{\gamma}$ is not normal ordered, indeed, the continuum limit leads to single-body decay terms, as in Eq.~\eqref{eq:one_body_decay_continuum}, which clearly do not capture the nonlinear and correlated nature of the coagulation decay. For the branching process \eqref{eq:branching} we obtain in the continuum limit the jump operator:
\begin{equation}
 L^\beta(x) = \lim_{a\to 0}\frac{1}{\sqrt{a}}L_{aj}^{\beta \pm} = \pm\sqrt{\widetilde{\Gamma}_{\beta}}\psi^\dagger(x)\partial_x\psi^\dagger(x)\psi(x) +\mathcal{O}(a^3), \quad \mbox{with} \quad \widetilde{\Gamma}_{\beta}= a^4\Gamma_\beta.
\label{eq:branching_continuum}
\end{equation}
Both in the case of coagulation and branching, the dissipation strength $\widetilde{\Gamma}_{\gamma}$ and $\widetilde{\Gamma}_{\beta}$, respectively, has units $[\mbox{length}^4 \, \,  \mbox{time}^{-1}]$. In the next Section, we discuss the TGGE method to describe the dynamics of the Fermi gas in the regime where the dissipation in Eqs.~\eqref{eq:one_body_decay_continuum}-\eqref{eq:branching_continuum} is weak.

\section{Reaction-limited dynamics and TGGE}
\label{sec:III_GGE}
\begin{figure}[t]
     \centering
     \includegraphics[width=0.8\columnwidth]{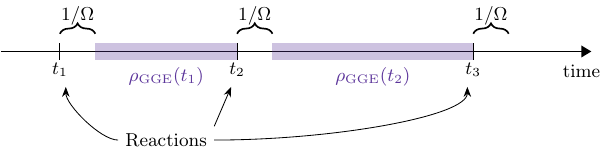}
     \caption{\textbf{Time dependent generalised Gibbs ensemble (tGGE).} Time evolution of the density matrix in the reaction-limited regime of weak dissipation $\widetilde{\Gamma}_{\nu} n^{z-2}/\Omega \ll 1$. Reactions occur on a much longer time-scale $\sim (n^z \widetilde{\Gamma}_{\nu})^{-1}$  than that of coherent motion $\sim n^{-2} \Omega^{-1}$, such that in-between consecutive reactions the system locally relaxes to the stationary state of the Hamiltonian. This local stationary state is a generalized Gibbs ensemble, which is for the here considered non-interacting Fermi gas a gaussian state. The time-dependence $\rho_{\mathrm{GGE}}(t)$ of the GGE accounts for the breaking of the conservation laws due to the reactions.}
     \label{fig:t-gge}
 \end{figure}

In the following we consider the RD dynamics of Sec.~\ref{sec:II_continuum} in the reaction-limited regime of weak dissipation $\widetilde{\Gamma}_{\nu} n^{z-2}/\Omega \ll 1$, with $\widetilde{\Gamma}_{\nu}=a^z \Gamma_{\nu}$, $n$ the particle density, while the exponent $z$ depends on the specific reaction $\nu$ considered. The reaction-limited regime can be analytically studied via the time-dependent generalized Gibbs ensemble (TGGE) \cite{TGGE1,TGGE2,TGGE3,TGGE4}, which we sketch in Fig.~\ref{fig:t-gge}. In particular, one exploits the separation of time scales between the dissipation, and coherent Hamiltonian evolution. The latter takes place on a much faster time scale than the former $n^{-2} \Omega^{-1} \ll (n^z\widetilde{\Gamma}_{\nu})^{-1}$, such that after each reaction the system relaxes locally to the stationary state $\rho_{\mathrm{SS}}(t)$ of the Hamiltonian evolution $[H,\rho_{\mathrm{SS}}(t)]=0$. Here, time dependence in the state $\rho_{\mathrm{SS}}(t)$ accounts for the fact that such state is stationary only for the Hamiltonian, but not for the full Lindbladian including dissipation \eqref{eq:dissipator_continuum}. At this point, one makes an assumption concerning the form of $\rho_{\mathrm{SS}}(t)$. Namely, one chooses it as a maximal entropy state fixed by the expectation values of all the (possibly multiple) extensive conservation laws $Q_m$ in the system: $\rho_{\mathrm{SS}}(t) \equiv \rho_{\mathrm{GGE}}(t)$
\begin{equation}
\rho_{\mathrm{GGE}}= \frac{e^{-\sum_m, \beta_m(t) Q_m}}{\mbox{Tr}[e^{-\sum_m \beta_m(t) Q_m}]},
\label{eq:gge_general}
\end{equation}
with $\{\beta_m(t)\}$ the set of Lagrange multipliers in one-to-one correspondence with the expectation values of the conserved charges in the initial state. This state is named generalized Gibbs ensemble \cite{GGE1,GGE2}. In the presence of dissipation it acquires a slow time dependence through the multipliers $\{\beta_m(t)\}$, hence the name TGGE. In formulas, the TGGE assumption then reads 
\begin{align}
    \lim_{\widetilde{\Gamma}\to 0}\rho(t=\tau/\widetilde{\Gamma}) = \rho_\mathrm{GGE}(\tau).
\label{eq:tgge_assumption_general}
\end{align}
We emphasize that Eq.~\eqref{eq:tgge_assumption_general} has to be understood in the local sense, i.e., it applies to local observables $\mathcal{O}$ in an infinite system. From the latter formula it is also clear that the TGGE describes the slow dynamics on the timescale $\widetilde{\Gamma}^{-1}$, with $t \to \infty$, $\widetilde{\Gamma}\to 0$, and $\tau =\widetilde{\Gamma} t$ held fixed.

In the specific case of quantum RD systems, the Hamiltonian in \eqref{eq:free_hamiltonian_continuum} is quadratic and it can diagonalized by Fourier transform. This allows to identify the conserved charges and the GGE state in a simple way. Our conventions for the Fourier transform are here reported as they will be needed later in Sec.~\ref{sec:IV_results}:
\begin{align}
    &\phi(k) =  \frac{1}{\sqrt{L}}\int_0^L \mbox{d}x \psi(x) e^{-ikx}, &\psi(x) =  \frac{1}{\sqrt{L}}\sum_{k\in\Lambda^*}\phi(k) e^{ikx}.    
\label{eq:fourier1}
\end{align}
In a gas in a finite volume and periodic boundary conditions the set $\Lambda^*$ of allowed momenta is $k_n= 2\pi n/L$, with $n=0, \pm 1 ,\pm 2, \dots \mathbb{Z}$ an arbitrary integer number. As the volume $L\to \infty$, momenta becomes dense on the real line $\Lambda^{\star} \equiv \mathbb{R}$ and one has
\begin{align}
  \tilde\phi(k) &= \frac{1}{\sqrt{2\pi}}\int_{\mathbb R}\mbox{d}x \ \psi(x)e^{-ikx}, \quad  \psi(x) =
  \frac{1}{\sqrt{2\pi}}\int_{\mathbb R} \mbox{d}k\ \tilde\phi(k) e^{ikx},
\label{eq:fourier2}
\end{align}
with $ \tilde\phi^{(\dagger)}(k) := \sqrt{\frac{L}{2\pi}}\phi^{(\dagger)}(k)$. We therefore see that $\tilde{\phi}(k)$ has units of $[\mbox{length}^{1/2}]$. From the Fourier mode operators \eqref{eq:fourier1} and \eqref{eq:fourier2} one can then define a ``scattering basis'' \cite{GGE1,GGE2,Doyon20} for the conserved charges $Q_m$ in terms of the Fourier occupation number operator $n_k= \phi^{\dagger}(k) \phi(k)$. The latter is not an extensive conserved charge, but it allows to express the extensive charges $\{Q_m\}$ as a linear combination of $\{n_k\}$ as 
\begin{equation}
Q_m = \sum_{k \in \Lambda^{\star}} f_m(k) n_k.
\label{eq:charges_asymptotic_basis}
\end{equation}
The function $f_m(k)$ is arbitrary, an example is $f_m(k)=k^m$, $m\in \mathbb{N}$, which gives an hermitean set of local conserved charges. From Eq.~\eqref{eq:charges_asymptotic_basis} one can then write the TGGE \eqref{eq:gge_general} as 
\begin{equation}
\label{eq:tgge}
    \rho_\mathrm{GGE}(t) = \frac{e^{-\sum_k \lambda_k(t) n_k}}{\mbox{Tr}[ e^{-\sum_k\lambda_k(t) n_k}]},
\end{equation}
with $\lambda_k(t)=\sum_{m} \beta_m(t) f_m(k)$. We note that for the calculations is neither necessary to pick a specific basis of charges $\{f_m(k)\}$ nor to decompose $\lambda_k$ in terms of the $\{\beta_m\}$. As a matter a matter of fact $\lambda_k$ fixes uniquely the GGE state and it is in one-to-one correspondence with the occupation function $C_k=\braket{\phi_k^{\dagger} \phi_k}$ in momentum space as 
 \begin{align}
 \left\langle \phi^\dagger_k\phi_p \right\rangle_\mathrm{GGE}(t) = Tr\left\lbrace  \phi^\dagger_k\phi_p \rho_\mathrm{GGE}(t) \right\rbrace = \frac{\delta_{k,p}}{1+e^{\lambda_k(t)}}=C_k \delta_{k,p}.
\label{eq:occupation_function}
\end{align}
In the limit $L \to \infty$, from \eqref{eq:fourier2}, we obtain
\begin{align}
 \left\langle \phi^\dagger_k\phi_p \right\rangle_\mathrm{GGE}(t) =: C_k(t)\delta_{k,p} \xrightarrow{L\to\infty}  \left\langle \tilde  \phi^\dagger_k\tilde \phi_p \right\rangle_\mathrm{GGE}(t) = C_k(t)\delta(k-p).
\label{eq:occupation_function_L_infty}
\end{align}
We note that for the GGE state \eqref{eq:tgge} to be physically meaningful, the function $\lambda_k$ must grow sufficiently fast as $|k|\to \infty$. In particular, we note that states with a flat occupation in momentum space $\lambda_k \equiv \lambda$, independent on $k$, which are possible on the lattice \cite{perfetto22,perfetto23,riggio2023}, do not correspond to a physically meaningful GGE (all charges expectation values would be divergent). To completely fix the dynamics of the TGGE state \eqref{eq:tgge}, one therefore needs an equation for the two-point function $C_q$. All the higher point functions are obtained from $C_q$ by applying Wick's theorem since the GGE state \eqref{eq:tgge} is gaussian for free fermions. The equation for $C_q$ is simply obtained from the Lindblad master equation using that $[H,n_q]=0$. In the lattice case, one has \cite{lossth1,lossth3,Rosso2022,lossth6,perfetto22,perfetto23,riggio2023} 
\begin{equation}
    \frac{dC_q(t)}{dt} = \sum_{j,\nu}\left\langle L_j^{\nu\dagger}\left[n_q, L_j^\nu\right]   \right\rangle_\mathrm{GGE}(t).
\label{eq:tGGE-rate_lattice}
\end{equation}
Taking the continuum limit, one obtains 
 \begin{equation}
 \label{eq:tGGE-rate}
    \frac{dC_q}{dt} =  \sum_{\nu}   \int_0^L dx \left\langle L^{\nu\dagger} (x)[ n(q), L^{\nu} (x) ] \right\rangle_\mathrm{GGE} (t),
\end{equation}
with the continuum limit of the jump operators $L^{\nu}(x)$ defined in Eqs.~\eqref{eq:dissipator_continuum} and \eqref{eq:jumps_continuum}. Here, the average $\braket{\dots}_{\mathrm{GGE}}$ is taken over the GGE state \eqref{eq:tgge}. Applying Wick's theorem, Eq.~\eqref{eq:tGGE-rate} then yields a closed differential equation for $C_q$. The mean particle density $n(t)$ is obtained from the latter via summation, respectively integration as $L\to \infty$, over all modes
\begin{align}
    n(t) = \frac{1}{L}\sum_{q\in \Lambda^{\star}} C_q(t) \xrightarrow{L\to\infty} \frac{1}{2\pi} \int_{\Lambda^*} dq\ C_q(t),
\label{eq:density_gge}
\end{align}
We emphasize here that the GGE state \eqref{eq:gge_general} describes relaxation of local observables in the thermodynamic limit, both for lattice and continuum systems. Equations \eqref{eq:tGGE-rate_lattice} and \eqref{eq:tGGE-rate} therefore apply only to the case of local jump operators. On the lattice locality means that  $L_j^{\nu}$ acts on a finite number of sites separated by a finite distance in the thermodynamic limit $L\to \infty$. All the jump operators in Eqs.~\eqref{eq:death}-\eqref{eq:branching} are local according to the previous definition. In the continuum limit of Eq.~\eqref{eq:tGGE-rate}, locality of the jump operators $L^{\nu}(x)$ implies that the latter is a pointwise function of the field $\psi(x)$ and its derivatives. This applies to all the jump operators we introduced in Eqs.~\eqref{eq:one_body_decay_continuum}-\eqref{eq:branching_continuum}. Consequently, we expect the TGGE rate equation \eqref{eq:tGGE-rate_lattice} and \eqref{eq:tGGE-rate} to correctly describe the reaction-limited dynamics for all the RD processes considered in the manuscript. For nonlocal reaction processes, such as long-range binary annihilation of pairs of particles, as discussed in Ref.~\cite{despres2025dynamics}, we, instead, expect the TGGE ansatz not to capture the associated weakly-dissipative dynamics.

It is also important to note that Eq.~\eqref{eq:tGGE-rate} describes the dynamics ensuing from homogeneous initial states. The generalization to inhomogeneous cases can be, however, performed by assuming that local relaxation to a GGE takes place pointwise in space. The GGE and the associated occupation function $C_q(x,t)$ therefore acquires an additional dependence on the space coordinate $x$. The left hand side of \eqref{eq:tGGE-rate} then acquires an additional contribution describing ballistic transport of the particles, while the left hand side takes the same form as in Eq.~\eqref{eq:tGGE-rate}. We will only consider homogeneous initial conditions in the whole manuscript for the sake of illustration purposes. A discussion of the extension of Eq.~\eqref{eq:tGGE-rate} to inhomogeneous cases for binary annihilation has been performed in Refs.~\cite{Rosso2022,gerbino2023largescale,gerbino2024kinetics,maki2024loss}.

In order to numerically quantify the exponent of the power law $n(\tau) \sim t^{\delta}$, it is useful to introduce the effective exponent $\delta(\tau)$ as 
\begin{align}
\label{def:delta}
    \delta(\tau):= -\frac{\ln(n(\tau)/ n(b\tau))}{\ln b},
\end{align}
where $b$ is a scaling parameter. If one has an asymptotic power law for the density $n(\tau)\sim\tau^{\delta}$, then in the limit of $\tau\to\infty$, $\delta(\tau)$ converges to the time-independent exponent $\delta$ independently of the chosen value for $b$. In the next Section, we specialize Eq.~\eqref{eq:tGGE-rate} to the various dissipative reaction processes introduced in Subsec.~\ref{subsec:continuum_gas} for the Fermi gas.

\section{Results}
\label{sec:IV_results}
This section contains the original results of the manuscript obtained from the TGGE equation \eqref{eq:tGGE-rate}. In Subsec.~\ref{subsec:2bresults}, we discuss the case of binary annihilation $2A\to\emptyset$ \eqref{eq:binary_ann_continuum_FB}. In Subsec.~\ref{subsec:3bresults}, we present the results for the reaction $3A\to \emptyset$ \eqref{eq:3ann_continuum}. In Subsec.~\ref{subsec:coagulation}, we present the results for coagulation $2A\to A$ \eqref{eq:coagulation_continuum}. In Subsec.~\ref{subsec:branching}, we eventually consider particle creating process, namely branching \eqref{eq:branching_continuum}, and study the stationary state phase transition obtained from the competition between branching, single-body decay \eqref{eq:one_body_decay_continuum} and binary annihilation.

\subsection{\textbf{Binary annihilation}}
\label{subsec:2bresults}
We insert the expression of $L^{2\alpha}(x)$ in Eq.~\eqref{eq:tGGE-rate} and use the Fourier transform to obtain \eqref{eq:fourier1}
\begin{align}
     \frac{dC_q}{dt} &=  \int_0^L dx \left\langle L^{2\alpha\dagger}(x) [ n(q), L^{2\alpha} (x) ] \right\rangle_\mathrm{GGE} (t) \nonumber \\
     &= \frac{\Tilde{\Gamma}_{2\alpha}}{L^2} \sum_{k,k'p,p'} \int_0^L dx \left\langle e^{-i(k+k')x} \phi^\dagger(k')( -ik')\phi^\dagger(k) \left[ n(q),e^{i(p+p')x} \phi(p)( ip')\phi(p') \right] \right\rangle_\mathrm{GGE} \nonumber \\
     &= \frac{\Tilde{\Gamma}_{2\alpha}}{L} \sum_{k,k'p,p'} \left\langle \phi^\dagger(k')( -ik')\phi^\dagger(k)    \left[ n(q), \phi(p)( ip')\phi(p') \right] \right\rangle_\mathrm{GGE} \frac{1}{L}\int_0^L dx e^{-i(k+k')x}e^{i(p+p')x} \nonumber \\
     &= -\frac{\Tilde{\Gamma}_{2\alpha}}{L} \sum_{k\in\Lambda^{\star}} (q-k)^2C_kC_q.
\label{eq:rate_alpha_intermediate}
\end{align}
From the third to the fourth line we used the commutation relation 
$[n(q),\phi(p),\phi(p')]=-\phi(p)\phi(p')(\delta_{p,q} +\delta_{p',q})$ and the Fourier representation of the Kronecker delta function
\begin{equation}
\delta_{k+k',p+p'} =   \frac{1}{L}\int_0^L dx e^{-i(k+k')x}e^{i(p+p')x}.
\label{eq:fouerier_Kronecker}
\end{equation}
In addition, Wick's theorem is applied in decomposing the four-point fermionic correlation function
\begin{equation}
\left\langle \phi^\dagger(k')\phi^\dagger(k)\phi(p)\phi(p') \right\rangle_\mathrm{GGE} = C_kC_{k'}(\delta_{k,p}\delta_{k',p'}-\delta_{k',p}\delta_{k,p'}).
\label{eq:wick_theorem}
\end{equation}
Taking now the limit $L\rightarrow \infty$ and introducing the rescaled time $\tau := t\Tilde{\Gamma}_{2\alpha}/(2\pi)$ one has
\begin{equation}
\label{eq:rate_alpha}
    \frac{dC_q(\tau)}{d\tau} = -\int_{-\infty}^\infty dk (q-k)^2C_k(\tau)C_q(\tau).
\end{equation}
Two important observations are here in order. First, in taking the infinite volume limit $L\to \infty$ in passing from Eq.~\eqref{eq:rate_alpha_intermediate} to \eqref{eq:rate_alpha}, we have extended the set $\Lambda^{\star}$ of allowed momenta to the whole real line $k \in \mathbb{R}$. This is possible since for binary annihilation there are no UV-divergences here, i.e., all the integrals in Eq.~\eqref{eq:rate_alpha} of $C_q$ are convergent at large $q$ values [assuming that the momentum occupation function vanishes for high momenta]. This is a consequence of the fact that the four point function \eqref{eq:wick_theorem} is normal ordered. Second, the same equation \eqref{eq:tGGE-rate} describes both the continuum limit of Eq.~\eqref{eq:2annihilation} and \eqref{eq:annihilation_interference}. This implies that physically different decay processes on the lattice yield the same decay when the continuum limit is taken. Equation \eqref{eq:rate_alpha} has been first derived in Ref.~\cite{Rosso2022} by taking the continuum limit $a\to 0$ of the TGGE equation obtained for the lattice process \eqref{eq:annihilation_interference} at $\theta=\pi/4$. In Ref.~\cite{gerbino2023largescale}, Eq.~\eqref{eq:rate_alpha} has been, instead, derived and extended to generic dimensions $d$ using the Keldysh action associated to the continuum limit of \eqref{eq:2annihilation}. The present derivation thereby shows in a unifying way from the TGGE how the same behavior arises both for \eqref{eq:2annihilation} and \eqref{eq:annihilation_interference} in the continuum limit.   
\begin{figure}[h]
    \centering
\includegraphics[width=0.495\columnwidth]{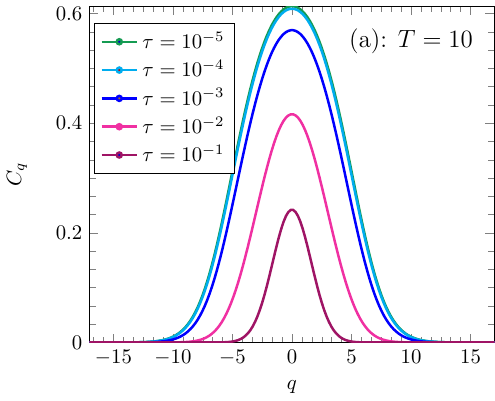}
\includegraphics[width=0.495\columnwidth]{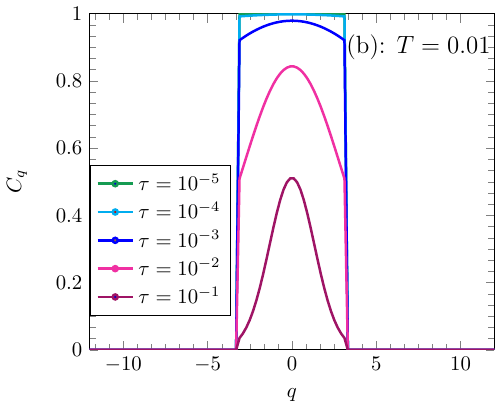}
    \caption{\textbf{Momentum occupation function at various times $\tau$ for binary annihilation.} (a) $C_q$ is plotted as a function of $q$ for increasing $\tau$ (top to bottom) for a thermal initial state with $T=10$ and initial density $n_0=1$. In panel (b) we plot the same times for a different temperature $T=0.01$ and the same $n_0$ value. The curves correspond to the numeric solution of Eq.~\eqref{eq:rate_alpha}. The occupation function becomes gaussian in time with a peak at $q=0$.}
    \label{fig:cq-alpha}
\end{figure}

\begin{figure}[h]
\includegraphics[width=0.495\columnwidth]{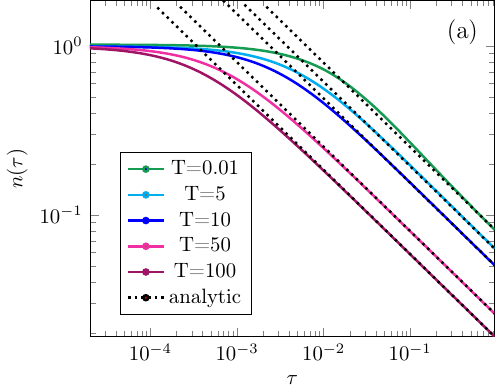}
\includegraphics[width=0.495\columnwidth]{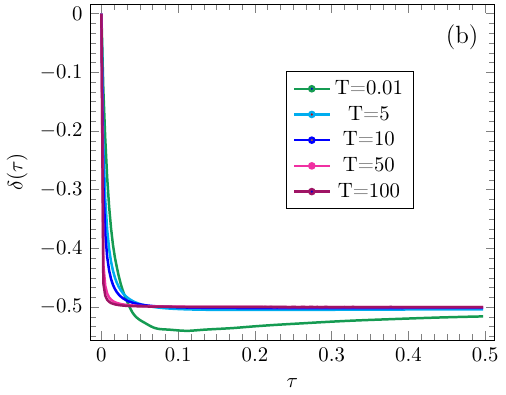}
  \caption{\textbf{Reaction-limited binary annihilation in the continuum limit.} (a) Log-log plot of the decay of the density $n(\tau)$ as a function of the rescaled time $\tau$ for increasing values of the temperature $T$ (from top to bottom).The solution of Eq.~\eqref{eq:rate_alpha} is  found with fourth-order Runge-Kutta integrator with the parameters  $d\tau=10^{-5}$ (time step of the numerical integrator) $dk:=2\pi/L=0.1$ (momentum grid step), $a=0.1$ (lattice spacing setting the maximum wavevector $|k|<k_{M}=\pi/a$). For the analytic curves (black-dashed lines) we used  Eq.~\eqref{n_ann}, while the coloured curves are the numerical solution of Eq.~\eqref{eq:rate_alpha}. (b) Effective exponent $\delta(\tau)$, see Eq.~\eqref{def:delta}, as a function of $\tau$ with $b=2$. Increasing the temperature from $T=0.01$ (green) to $T=100$ (red) accelerates the convergence to the power law exponent $\delta=-1/2$. In both panels the initial density is set to $n_0=1$.}
  \label{fig:alpha}
\end{figure}

In Fig.~\ref{fig:cq-alpha}, we plot the occupation function $C_q(\tau)$ calculated by numerically solving Eq.~\eqref{eq:rate_alpha} for an initial thermal state
\begin{equation}
\label{eq:n_fd}
    C_q(0) = \dfrac{1}{1+e^{\beta(\frac{1}{2}q^2-\mu(\beta))}},
\end{equation}
with $\beta$ the inverse temperature and $\mu$ the chemical potential (fixing the initial density $n_0$). In Fig.~\ref{fig:cq-alpha}(a) we consider a large temperature, while in Fig.~\ref{fig:cq-alpha}(b) a low temperature is taken and the initial distribution is close to a Fermi-sea state (ground state of the fermionic Hamiltonian \eqref{eq:free_hamiltonian_continuum}). In both the cases, we see that high momentum modes decay faster to zero. Independent of the initial temperature, the occupation function $C_q$ attains a gaussian shape during the time evolution. In Fig.~\ref{fig:alpha}(a), we plot the associated decay of the density $n(\tau)$ \eqref{eq:density_gge} as a function of $\tau$ for various values of the temperature $T$. We see that upon increasing the temperature the amplitude of the decay is rescaled, while the exponent of the power-law $n(\tau)\sim t^{\delta}$ does not change. Namely, we find that $\delta=-1/2$ independently of the initial state, as shown in Fig.~\ref{fig:alpha}(b) from the plot of the effective exponent $\delta(\tau)$ \eqref{def:delta}. Upon increasing $T$, the effective exponent also converges faster to the value $\delta=-1/2$.

In order to quantify these features we study the $\tau\to\infty$ asymptotics of Eq.~\eqref{eq:rate_alpha}. We consider symmetric initial conditions $C_k(0) = C_{-k}(0)$ throughout, as in Eq.~\eqref{eq:n_fd}. Since this symmetry is conserved during the time evolution by Eq.~\eqref{eq:rate_alpha} one has that $C_k(\tau)=C_{-k}(\tau)$. For zero-temperature initial states, the asymptotics has been worked out in Refs.~\cite{Rosso2022,gerbino2023largescale}. Here, we extend the calculation to generic-finite temperatures initial states. In the following we set $n_0:=n(0)$ and $C_0:=C_0(0)$. From Eq.~\eqref{eq:rate_alpha}, we obtain
\begin{align}
\label{eq:rate-alpha-sym}
    \dot{C_q}(\tau) &= -\int_{-\infty}^\infty dk (q^2+k^2)C_k(\tau)C_q(\tau)= -C_q(\tau)\int_{-\infty}^\infty dk k^2C_k(\tau) -2\pi C_q(\tau)q^2 n(\tau),
\end{align}
which via time integration yields
\begin{equation}
\label{eq:alpha-c-formal}
C_q(\tau) = C_q(0)\sqrt{\frac{n(\tau)}{n_0}} e^{-2\pi q^2\nu( \tau)}, \quad \mbox{with} \quad \nu(\tau)=\int_{0}^{\tau}\mbox{d}\tau' n(\tau').
\end{equation}
This formal solution already shows that $C_q(\tau)$ develops a gaussian structure with a variance shrinking in time. The result in Eq.~\eqref{eq:alpha-c-formal} interestingly shows that at long times the fermionic gas asymptotically turns into a low-temperature classical gas described by a Maxwell-Boltzmann distribution. The associated temperature, $T(t)\sim \nu(t)^{-1}\sim t^{-1/2}$, is in fact lowered as time passes, as we will now show. From Eq.~\eqref{eq:alpha-c-formal}, we depart from the calculation of Refs.\cite{Rosso2022,gerbino2023largescale}, where the expression is initialized to a Fermi-sea step distribution. We do this by performing a saddle-point approximation, which is justified since the time-integrated density $\nu(\tau)$ diverges at long times:
\begin{align}
    \sqrt{n(\tau)n_0}  &\approx \frac{C_0}{2\pi}\int_{-\infty}^\infty dq e^{-2\pi q^2\nu( \tau)} = \dfrac{C_0}{2\pi\sqrt{2\nu(\tau)}}.
\end{align}
Using that $n(\tau)= \dot\nu(\tau)$ one finds
\begin{equation}
\label{n_ann}
      n(\tau) \sim \dfrac{C_0}{4\pi\sqrt{\tau n_0}}.
\end{equation}    
In order to make the temperature dependence explicit, we use the relation between the chemical potential $\mu$ and $n_0$ from Eq.~\eqref{eq:n_fd}
\begin{equation}
\label{eq:n0-field}
    n_0 =\int_{-\infty}^\infty \frac{dq}{2\pi} \dfrac{1}{1+e^{\beta(\frac{1}{2}q^2-\mu(\beta))}}= \frac{1}{\pi\sqrt{2\beta}}\int_{0}^\infty dz \dfrac{z^{-\frac{1}{2}}}{1+e^{z-\beta\mu(\beta)}}=  - \frac{1}{\sqrt{2\pi\beta}} \text{Li}_\frac{1}{2}\left(-e^{\beta\mu(\beta)}\right).  
\end{equation}
Here, $\text{Li}_s(z)$ is the polylogarithm \cite{NIST:DLMF}. The numerical solution of the latter shows that $\beta\mu<0$ for high temperatures $T$ and therefore $ e^{\beta\mu}\ll 1$.
We then find
\begin{align}
     n_0 \approx \frac{1}{\sqrt{2\pi\beta}}e^{\beta\mu(\beta)}
 \quad\Rightarrow \quad \mu(\beta)=\frac{1}{2\beta}\ln(2\pi\beta n_0^2). 
\end{align}
which we now insert in Eq.~\eqref{n_ann}:
\begin{align}
    n(\tau) = \dfrac{C_0}{4\pi\sqrt{\tau n_0}}= \dfrac{1}{4\pi\sqrt{\tau n_0}\left(1+e^{-\beta\mu}\right)}\approx  \dfrac{e^{\beta\mu}}{4\pi\sqrt{\tau n_0}}=\sqrt{\frac{\beta n_0}{8\pi\tau}} \sim \left(\tilde\Gamma T t\right)^{-\frac{1}{2}}. \label{eq:n_th_a}
\end{align}
Hence we can see that the amplitude of the decay is rescaled according to the temperature. As the temperature is increased, the decay becomes faster, but its asymptotic exponent is not changed. When comparing Eq.~\eqref{eq:n_th_a} with the exact numerical solution of \eqref{eq:rate_alpha} perfect agreement is found at long times already for moderately low values of $T$. This result shows that decay exponents in quantum RD systems that differ from mean-field predictions are not a low-temperature effect, as their values are robust with respect to considering thermal fluctuations in the initial state. This result is based on particle repulsion due to the fermionic hard-core constraint. In quantum reaction-limited RD, particle anticorrelations are therefore rooted in the quantum statistics and not in the appearance of depletion zones at long times. The latter are, indeed, by construction absent in the reaction-limited regime due to the fast coherent mixing. The mechanism behind the emergence of the universal quantum reaction-limited decay \eqref{eq:n_th_a} is therefore fundamentally different from that underlying the appearance of decay different from mean field in classical diffusion-limited RD. Furthermore, the accelerated decay \eqref{eq:n_th_a} comes from the fact that the quasiparticle energy spectrum is unbounded in continuum space. Consequently at high temperatures higher and higher momenta are thermally excited (conversely fewer and fewer modes around $q=0$ are excited). These momenta decay rapidly.   

In a lattice, on the contrary, this is not possible since the quasiparticle energy spectrum is bounded, as we are now going to show. On the lattice, the TGGE equation for the reaction \eqref{eq:annihilation_interference} has been worked out in Ref.~\cite{lossth1} for $\theta=\pi/4$ and in Ref.~\cite{perfetto22} for generic theta values:
\begin{align}
\label{eq:rate-l}
    \frac{d C_q(\tau)}{d\tau}  = -\frac{1}{N} \sum_k g_\theta(k,q) C_k C_q
\end{align}
with $g_\theta(k,q) = 2(1-\cos(k-q)) +\sin(2\theta)(2\cos(k+q) - \cos(2k) - \cos(2q))$. In contrast to the continuum, we need to treat distinct values of $\theta$ separately.  
\begin{figure}[h]
    \centering
   \includegraphics[width=0.495\columnwidth]{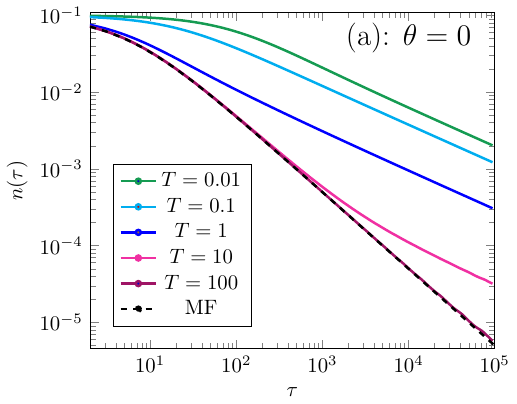}
    \includegraphics[width=0.495\columnwidth]{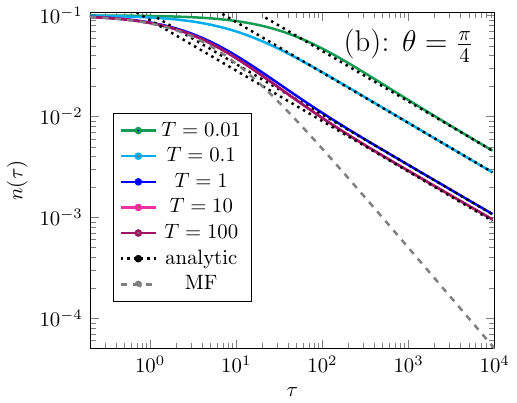}
    \caption{\textbf{Reaction-limited binary annihilation on the lattice}. Log-log plot of the decay of the density $n(\tau)$ as a function of the rescaled time $\tau$ for increasing values of the temperature (from top to bottom) and initial filling $n_0=0.1$. (a) Numerical solution of Eq.~\eqref{eq:rate-l} for $\theta=0$ with  parameters $a=0.1$, $dk=0.1$. (b) Numerical solution of Eq.~\eqref{eq:rate-l} for $\theta=\pi/4$. The dashed-dotted lines represent the asymptotic behavior at long times computed from Eqs.~\eqref{eq:initial_filling_lattice} and~\eqref{eq:asymp_lattice}. For high (bottom-purple line) and low temperature (top green line) the asymptotics approaches \eqref{eq:l_as_th} and \eqref{eq:l_as_0}, respectively.
\label{fig:lattice_binary_ann}
    }
\end{figure}

The values $\theta=0,\pi$ corresponds to the classical binary annihilation jump operator \eqref{eq:2annihilation}. For this process, it was shown in Refs.~\cite{perfetto22,riggio2023} the asymptotic decay $n(\tau) \sim \tau^{-1/2}$ as long as initial states featuring quantum coherences in real spaces are considered, i.e., an initial distribution $C_q(0)$ not flat in $q$ is considered. In Fig.~\ref{fig:lattice_binary_ann}(a), we additional investigate the role of a finite temperature $T$ in the initial state. The initial occupation function $C_q(0)$ is not flat in momentum
\begin{align}
\label{eq:initial_filling_lattice}
C_q(0) = \frac{1}{1+e^{-\beta \cos q-\beta\mu}}, \quad \Rightarrow \quad n_0 =\int_{-\pi}^{\pi}\frac{d q}{2\pi} C_q(0).
\end{align}
The decay exponent is then $\delta=-1/2$ according to the previous discussion. Importantly, upon increasing the temperature the time window where mean-field decay applies extends. For very high $T$ it is now not possible to excite high momenta, since $q\in [-\pi,\pi]$ on the lattice. As a consequence, the initial distribution \eqref{eq:initial_filling_lattice} becomes asymptotically $C_q(0)=1/2$ and mean field is recovered. This is the fundamental difference compared to the continuum description depicted in Fig.~\ref{fig:alpha}. We reiterate that a flat occupation function $C_q(0)=\mbox{const}$ is not physically meaningful in the continuum since $C_q$ must be decreasing rapidly enough at large $q$ to ensure convergence of the charges expectation values, as commented after Eq.~\eqref{eq:occupation_function_L_infty}. Mean-field decay in the reaction-limited regime of binary annihilation is therefore a lattice effect absent in the associated continuum description, as shown in Fig.~\ref{fig:alpha}. 

Finally, we study the case $\theta=\pi/4$ in order to understand the effect of the simultaneous presence of coherences, introduced by annihilation \eqref{eq:annihilation_interference} reaction on the lattice, and temperature onto the density decay. This leads to dynamical behavior that is qualitatively different from the $\theta=0$ value. This is shown in Fig.~\ref{fig:lattice_binary_ann}(b). For zero-temperature initial states the same power-law exponent $\delta=-1/2$ is found \cite{lossth1,perfetto22}. Upon increasing $T$, however, the decay of $n(\tau)$ versus $\tau$ does not approach the mean-field power-law. To quantify this effect we extend the asymptotics derived in Ref.~\cite{lossth1} to arbitrary initial states [still keeping the condition $C_{q}(0)=C_{-q}(0)$]. The first steps are analogous to those we performed to get to \eqref{eq:alpha-c-formal}. In particular, starting from \eqref{eq:rate-l} with $\theta=\pi/4$ and performing the thermodynamic limit and using the normalization \eqref{eq:density_gge} yields an integrable differential equation with a formal exponential solution:
\begin{align}
    C_q(\tau) = \frac{C_q(0)}{2\pi}\sqrt{\frac{n(\tau)}{n_0}}e^{-2 \nu(\tau)\sin^2q}.
\end{align}
Again we integrate over all modes and use the saddle-point approximation and thereby extending the integration domain to $k \in (-\infty,\infty)$:
\begin{align}
    \sqrt{n(\tau)n_0} &= \frac{1}{2\pi}\int_{-\pi}^\pi dq\ C_q(0)e^{-2 \nu(\tau) \sin^2q}\approx \frac{1}{2\pi}\left(C_0+\frac{1}{2}C_{-\pi}+\frac{1}{2}C_{\pi}\right)\int_{-\infty}^\infty dq\ e^{-2 \nu(\tau) q^2}.
\end{align}
The three terms on the right hand side arise from the three minima of the function $\sin^2 q$ in the interval $[-\pi, \pi]$. Two minima $q_{1,2}= \pm \pi$ lie exactly at the boundaries of the integration interval and therefore contribute with a multiplicative factor of $1/2$. Thus we have ($C_{-\pi}=C_{\pi}$)
\begin{align}
    \sqrt{n(\tau)n_0}
    &=\frac{C_0+C_{\pi}}{\sqrt{8\pi \nu(\tau)}}  \quad\Rightarrow \quad n(\tau)=\frac{C_0+C_{\pi}}{4\sqrt{\pi \tau n_0}}.
\label{eq:asymp_lattice}
\end{align}
To fix the temperature dependence inside the factors $C_0$ and $C_{\pi}$, we again link the chemical potential $\mu$ from the initial occupation $n_0$. In particular, for high temperature, i.e., $\beta\rightarrow 0 \Rightarrow $, the product $\beta\mu$ approaches a constant and $\beta\cos q\rightarrow 0$ since the energy spectrum is bounded on the lattice. From this, as shown in \ref{app:Appendix1}, one eventually finds the density asymptotics 
\begin{equation}
\label{eq:l_as_th}
    n(\tau)|_{\beta\to0} = \lim_{\beta\to0}\frac{\sqrt{n_0}}{2\sqrt{\pi\tau}(1-\beta^2(1-n_0)^2)}=\frac{\sqrt{n_0}}{2\sqrt{\pi\tau}}.
\end{equation}
Conversely, in the zero temperature limit, $\beta\to\infty$, the initial distribution is a Fermi sea with $C_0=1$ and $C_\pi=0$ (as long as $n_0<1$), hence
\begin{equation}
\label{eq:l_as_0}
    n(\tau)|_{\beta=\infty} = \frac{1}{4\sqrt{\pi n_0\tau}}.
\end{equation}
Equation \eqref{eq:l_as_0} is consistent with the asymptotics derived in Ref.~\cite{lossth1} for the density $n(\tau)$ of the Fermi gas at long times $\tau$ in a zero-temperature initial state and small initial filling $n_0$. In the present work, Eq.~\eqref{eq:l_as_0} is derived for generic values of $n_0$ as the limiting case of Eq.~\eqref{eq:asymp_lattice}, which is valid for arbitrary temperature $T$. The decay amplitude is therefore bounded between the curves \eqref{eq:l_as_th} in the limit $\beta\to0$ ($T\to\infty$) and \eqref{eq:l_as_0} at zero temperature as can be observed in Fig.~\ref{fig:lattice_binary_ann}(b). From here one also notes that for $\theta=\pi/4$ and initial filling $n_0>0.5$, then $ n(\tau)|_{\beta=0} > n(\tau)|_{\beta=\infty}$, such that increasing the temperature slows down the decay. This is different from the case $\theta=0$ where upon increasing the temperature the density decay accelerates irrespectively of the value of the initial filling $n_0$. The result in \eqref{eq:l_as_th} is also different from the continuum case \eqref{eq:n_th_a} since the decay amplitude is bounded as $T$ increases. This is again an effect due to the bounded dispersion relation on the lattice.

In concluding this section we mention that the jump operator \eqref{eq:binary_ann_continuum_FB} describes the continuum limit of \eqref{eq:annihilation_interference} for all values of $\theta$ but the special values $\theta=\pi(4n-1)/4$, with $n\in \mathbb{Z}$. In these cases the rate $\widetilde{\Gamma}_{2\alpha}$ vanishes and one needs to consider the next-order term in the expansion. We, however, checked (we report the details in \ref{app:Appendix1} for the sake of brevity) that the asymptotic decay of the particle density in the gas does not change. In particular, one still has $n(\tau) \sim \tau^{-1/2}$, with the amplitude of the decay being rescaled by the temperature of the initial state.

\subsection{\textbf{Three-body annihilation}}
\label{subsec:3bresults}
The calculation for three-body annihilation is quite similar to that for the binary case. It amounts to inserting the Fourier transform \eqref{eq:fourier1} of $L^{3\alpha}$ \eqref{eq:3ann_continuum} into Eq.~\eqref{eq:tGGE-rate}. For $3A\to \emptyset$ one then needs to evaluate normal-ordered six-point fermionic correlation functions. This is achieved by Wick's theorem with similar, but longer, calculations as in Eq.~\eqref{eq:wick_theorem}. We do not report the intermediate steps for the sake of brevity, while we write the final kinetic equation for the two-point function $C_q$:  
\begin{equation}
    \frac{\mbox{d}C_q(\tau)}{\mbox{d}t} = -\frac{1}{18}\int_{-\infty}^{\infty} dk \int_{-\infty}^{\infty} dp \, C_q C_p C_k \, \left[(k-q)^3 -(k-p)^3 + (q-p)^3 \right]^2,
\label{eq:3ann_kin_eq}
\end{equation}
with the rescaled time $\tau=\widetilde{\Gamma}_{3\alpha}t/(2\pi)^2$ and the limit $L\to \infty$ already taken. As in the case of binary annihilation, we can safely extend both the $k$ and $p$ integrals of \eqref{eq:3ann_kin_eq} to the whole real line. This again originates from the fact that the jump operator \eqref{eq:3ann_continuum} is already normal ordered. The corresponding six-point fermionic correlation is consequently also normal ordered and no UV-divergences appear. Equation \eqref{eq:3ann_kin_eq} coincides with the $a\to 0$ continuum limit of the TGGE equation derived in Ref.~\cite{perfetto23} for the lattice formulation \eqref{eq:3annihilation} of the three-body annihilation process. We discuss in the following the case of three-body losses with a thermal initial occupation function $C_q(0)$ as in Eq.~\eqref{eq:n_fd}. 
\begin{figure}[h]
    \centering
\includegraphics[width=0.495\columnwidth]{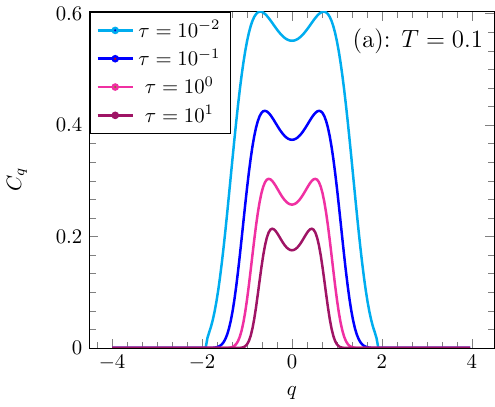}
\includegraphics[width=0.495\columnwidth]{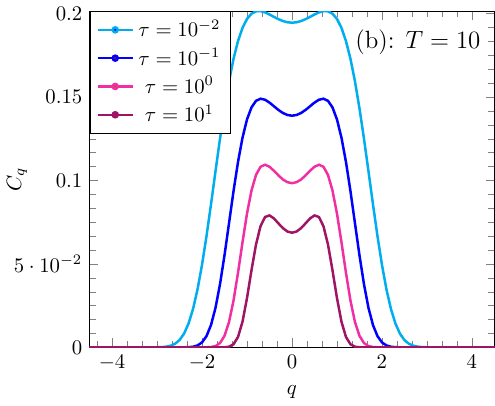}
    \caption{\textbf{Momentum occupation function for various values of time $\tau$ for three-body annihilation.} (a) $C_q$ is plotted as a function of $q$ for increasing $\tau$ (top to bottom) for a thermal initial state with $T=0.1$ and $n_0=1$. In panel (b) we plot the same times for a different temperature $T=10$ and the same $n_0=1$. The curves correspond to the numeric solution of Eq.~\eqref{eq:3ann_kin_eq}. The occupation function is non gaussian in time and it develops a double peak profile.}
    \label{fig:cq-alpha_3ann}
\end{figure}

Due to the more complicate structure of the right hand side of Eq.~\eqref{eq:3ann_kin_eq}, compared to the $2A\to \emptyset$ case \eqref{eq:rate_alpha}, we are not able to derive the explicit dependence of $C_q$ on $q$ as in Eq.~\eqref{eq:alpha-c-formal}. We can, however, understand the fundamental differences in the dynamics of $C_q$ compared to the binary annihilation case by numerically solving Eq.~\eqref{eq:3ann_kin_eq}. We show the corresponding results in Fig.~\ref{fig:cq-alpha_3ann}, where we plot the occupation function $C_q$ as a function of $q$ for various values of the rescaled time $\tau$. In Fig.~\ref{fig:cq-alpha_3ann}(a), we consider the case of an initial low-temperature state, while in Fig.~\ref{fig:cq-alpha_3ann}(b) a high-temperature initial distribution $C_q(0)$ is taken. In both the cases, we see that the distribution develops a double-peak profile with two local maxima at $q=\pm q_0$ with $q_0\neq 0$. Apparently for the process $3A\to \emptyset$ the momentum occupation function $C_q(\tau)$ never approaches a gaussian form even when $C_q(0)$ is peaked around $q=0$. This result is in stark contrast with the case of binary annihilation discussed in Eq.~\eqref{eq:alpha-c-formal}, and it shows that for three-body annihilation the occupation function in momentum space $C_q(\tau)$ never assumes a Maxwell-Boltzmann distribution. The fermionic gas subject to weak three-body annihilation is therefore never described by the statistical distribution of a classical low-temperature gas. As the temperature $T$ increases, cf. Fig.~\ref{fig:cq-alpha_3ann}(a) and \ref{fig:cq-alpha_3ann}(b), the double-peak shape of $C_q$ does not change, while the decay of $C_q$ accelerates. This is caused by the fact that at higher temperatures, higher momenta are initially excited and these momenta decay faster, as commented also for the process $2A\to\emptyset$ in Subsec.~\ref{subsec:2bresults}. 

\begin{figure}[t]
\includegraphics[width=0.495\columnwidth]{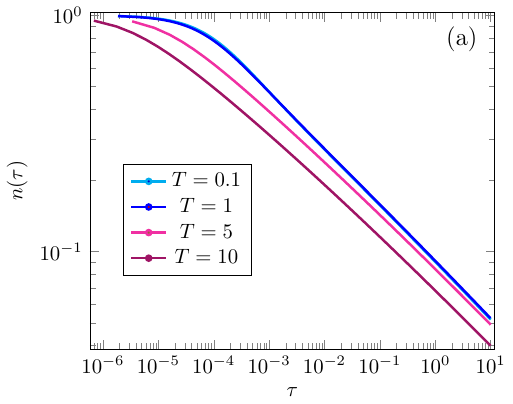}
\includegraphics[width=0.495\columnwidth]{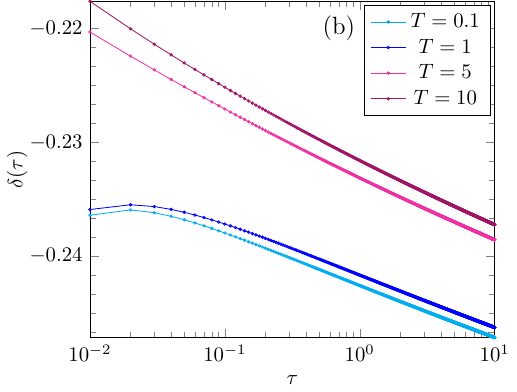}
  \caption{\textbf{Reaction-limited three-body annihilation in the continuum limit.} (a) Log-log plot of the decay of the density $n(\tau)$ as a function of the rescaled time $\tau$ for increasing values of the temperature $T$ (from top to bottom).The solution of Eq.~\eqref{eq:3ann_kin_eq} is found with Runge-Kutta four method  with $L=20 000$ values for $k$ ($dk:=2\pi/L$ is the spacing of the momentum grid), while the maximum wavevector $k_{M}$ is equal $k_M=6$ for $T=0.1,1$ and $k_M=12$ for $T=5,10$. (b) Effective exponent $\delta(\tau)$ \eqref{def:delta} as a function of $\tau$, with $b=2$. Increasing the temperature (from bottom to top) the absolute value $|\delta(\tau)|$ of the effective exponent decreases. The latter for all temperatures monotonically decreases as a function of $\tau$ indicating a non-algebraic asymptotic decay of the particle density. In both panels the initial density is $n_0=1$.}
  \label{fig:alpha_3ann}
\end{figure}

Accordingly also the particle density $n(\tau)$ decays faster in time $\tau$ upon increasing the temperature. This is shown in Fig.~\ref{fig:alpha_3ann}(a). The fundamental difference, however, compared to the binary annihilation case in Eq.~\eqref{eq:n_th_a} and Fig.~\ref{fig:alpha} is that the ensuing decay law is not algebraic for every value of the temperature $T$. This is shown in Fig.~\ref{fig:alpha_3ann}(b), where we plot the effective exponent $\delta(\tau)$ as a function of $\tau \leq 10$ for the same values of $T$ used in panel (a). We see that in the time window shown $\delta(\tau)$ decreases without converging to any value, in stark contrast with the case of Fig.~\ref{fig:alpha}(b) for $2A \to \emptyset$ for which the effective exponent rapidly converges to $-1/2$.       
Non-algebraic decay for $n(\tau)$ has previously been observed in Ref.~\cite{perfetto23} for the lattice \eqref{eq:3annihilation} problem for zero-temperature $T=0$ states. In the lattice case, however an algebraic behavior with effective exponent approximately $\delta(\tau)\approx -0.25$ was also found for intermediate times $\tau \lesssim 10^5$, before the non-algebraic asymptotics kicking in. In the continuum case, shown in Fig.~\ref{fig:alpha_3ann}(b), instead, even at low temperatures no intermediate algebraic decay is present and the effective exponent shows monotonic decay already from very short time scales ($\tau \gtrsim 0.1$ for $T=0.1,1$). We also mention that corrections to algebraic scaling have been attained in Ref.~\cite{marche2024universality} for the problem of two spinless bosons subject to two-body losses. Therein, arbitrary large loss rates were considered, so the dynamics is not necessarily reaction-limited, and the mean number of particles showed on the lattice algebraic decay with superimposed logarithmic corrections. Such corrections disappear, however, when the continuum limit is taken. In our case, instead, the corrections to the algebraic decay of $n(\tau)$ apply in the reaction-limited regime. 

\begin{figure}[h]
    \centering
\includegraphics[width=0.5\columnwidth]{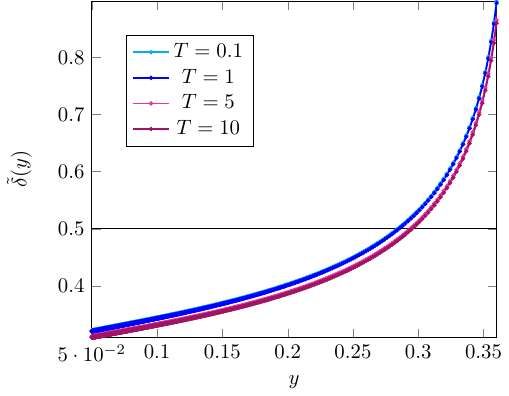}
    \caption{\textbf{Effective exponent for three-body annihilation in the continuum limit.} Plot of the effective exponent $\tilde{\delta}(y)$ in Eq.~\eqref{eq:tilde_delta} as a function of $y$ in Eq.~\eqref{eq:yvariable} for the fermionic gas in continuum space. The values of the temperature $T$ of the initial state and the other parameters are set as in Fig.~\ref{fig:alpha_3ann}. The effective exponent $\tilde{\delta}(y)$ does not converge for small $y$ (long times $\tau$). Namely, it attains values which are smaller than the saturation value $0.5$ (horizontal black line) obtained for classical-diffusion-limited three-body annihilation. This result shows that the non-algebraic decay observed for three-body annihilation in the Fermi gas does not feature logarithmic corrections.}
    \label{fig:delta_tilde}
\end{figure}

The analysis of $\delta(\tau)$ in Fig.~\ref{fig:alpha_3ann}(b), however, does not allow to conclude whether such corrections are logarithmic or not. In order to better understand this point we plot in Fig.~\ref{fig:delta_tilde} the effective exponent 
\begin{equation}
\tilde{\delta}(y)=-\frac{\log(\tilde{n}(y b)/\tilde{n}(y))}{\log b},
\label{eq:tilde_delta}
\end{equation}
as a function of
\begin{equation}
y=\frac{\log \tau}{\tau}.
\label{eq:yvariable}
\end{equation}
In the previous equation, we defined $\tilde{n}(y)=n(\tau(y))$, with $n$ the particle density. The function $y(\tau)$ in Eq.~\eqref{eq:yvariable} is, indeed, invertible for long times ($\tau>e$) and the inverse function $\tau(y)$ is well defined. We note that $y(\tau)$ is a decreasing function of $\tau$ for long times, such that small $y$ values correspond to large $\tau$ values. If the density $n(\tau)$ displays logarithmic corrections $n(\tau)\sim (\log \tau/\tau)^{\alpha}$ at long times, then $\tilde{n}(y) \sim y^{\alpha}$ and $\tilde{\delta}(y)$ converges to the exponent $\alpha$ for small $y$ values. In classical diffusion-limited three-body annihilation, $\alpha=1/2$ \cite{hinrichsen2000non,vladimir1997nonequilibrium,henkel2008non,tauber2014critical,Malte_Henkel_2004}. In Fig.~\ref{fig:delta_tilde}, on the contrary, we observe that $\tilde{\delta}(y)$ does not converge for small $y$ independently from the temperature $T$ of the initial state. We therefore conclude that the decay of the particle density for quantum reaction-limited three-body annihilation of the Fermi gas is not algebraic and it does not display logarithmic corrections. The markedly different behavior compared to the classical model is again caused by the different mechanism behind the decay in the two cases. In the classical diffusion-limited model, logarithmic corrections stem from spatial fluctuations, which are marginal in one dimension. In the quantum model, instead, fermionic statistics determine particle anticorrelations, even in the reaction-limited regime, causing the unexpected behavior of $\tilde{\delta}(y)$. 

We close this section commenting on the differences between the continuum and the lattice model. In the lattice case, one observes mean-field decay at intermediate times as the temperature $T$ is raised. This result is similar to that for $2A\to \emptyset$ in Fig.~\ref{fig:lattice_binary_ann}(a). In the case of $3A\to\emptyset$, mean-field decay is ruled given by the exponent $n(\tau)\sim \tau^{-1/2}$. In the continuum model, instead, mean-field decay is never observed regardless of the temperature, as commented above. The physics behind this difference between lattice and continuum is again based on the finite maximal velocity of particles on the lattice.

\subsection{\textbf{Coagulation}}
\label{subsec:coagulation}
\begin{figure}[h]
    \centering
    \includegraphics[width=0.495\columnwidth]{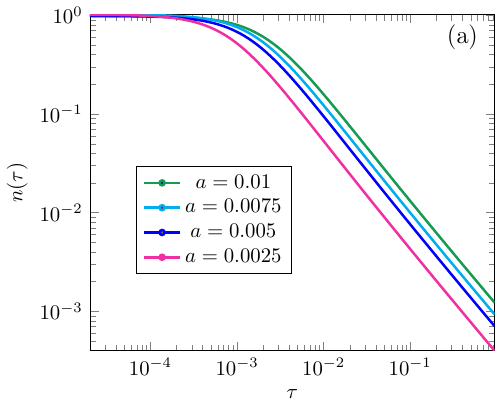}
    \includegraphics[width=0.495\columnwidth]{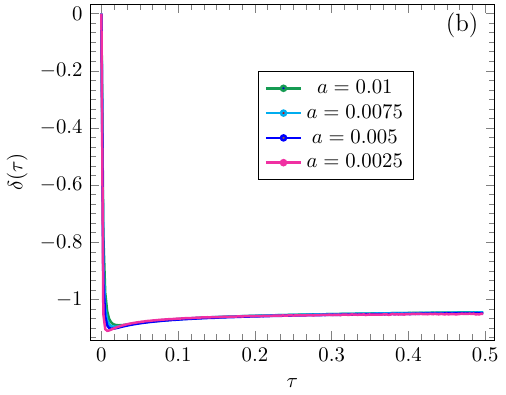}
    \caption{\textbf{Reaction-limited dynamics of coagulation in the continuum limit.} (a) Log-log plot of the particle density $n(\tau)$ as a function of the rescaled time $\tau$ for decreasing values of $a$ (from top to bottom). The initial state has density $n_0=1$ and temperature $T=0.01$. The parameter $a$ sets the maximum wavevector (UV cutoff) allowed $|k|<\pi/a$. The amplitude of the decay depends on the value of $a$. (b) Effective exponent $\delta(\tau)$ as a function of $\tau$ numerically obtained from Eq.~\eqref{eq:rate-gamma} for the same values of $a$, temperature and initial density. The parameters are $dk=0.1$ for the momentum grid. The effective exponent $\delta(\tau)$ converges to $-1$ independently on the value taken for $a$.}
    \label{fig:conv}
\end{figure}

Analogously to the previous discussed reactions, we insert the Fourier transform of Eq.~\eqref{eq:coagulation_continuum} into the TGGE rate equation \eqref{eq:tGGE-rate}. The ensuing fermionic six-point functions (in this case it is not normal ordered since $L^{\gamma \pm \dagger}(x) L^{\gamma \pm}(x)$ is not) is evaluated by Wick's theorem, as detailed in \ref{app:Appendix3}. This leads to the following equation of motion for the occupation function $C_q$ 
\begin{align}
\label{eq:gamma_rate-sum-0}
     \frac{dC_q}{dt}
     = -\frac{\widetilde{\Gamma}_\gamma}{L^2}\left[\sum_{k,p} (q-k)^2C_k C_q -\frac{1}{2}\sum_{k,p} (k-p)^2C_kC_p \right].
\end{align} 
The first term on the right hand side contains a free sum $\sum_{p}$ over all momenta. In this sum, one is not allowed to extend the upper summation limits to infinity, as we did in the previous cases of $k-$body annihilation. We therefore introduce a UV cutoff $|k| \leq k_M$, with $k_M=\pi/a$. In the thermodynamic limit $L\rightarrow\infty$ this implies constraining the integral upper/lower limits to a finite value
\begin{equation}
\label{eq:uv-cutoff}
    \frac{1}{L}\sum_p \rightarrow \frac{1}{2\pi}\int_{-\frac{\pi}{a}}^{\frac{\pi}{a}}dp.
\end{equation}
Taking the thermodynamic limit $L\to \infty$, with the regularized sum \eqref{eq:uv-cutoff} and introducing the rescaled time $\tau=\widetilde{\Gamma}_{\gamma}t/(2\pi)$, one has:
\begin{align}
\label{eq:gamma_rate-sum}
     \frac{dC_q}{dt}
     = -\widetilde{\Gamma}_\gamma\left[\frac{1}{a}\int_{-\pi/a}^{\pi/a} dk (q-k)^2C_k C_q -\frac{1}{2}\int_{-\pi/a}^{\pi/a}\frac{dk}{2\pi} \int_{-\pi/a}^{\pi/a} dp (k-p)^2 C_k C_p \right].
\end{align}
For thermal initial states of the form \eqref{eq:n_fd}, the symmetry $C_{k}(0)=C_{-k}(0)$ is preserved at all times $C_k(\tau)=C_{-k}(\tau)$ and one can further simplify Eq.~\eqref{eq:gamma_rate-sum} as
\begin{align}
\label{eq:rate-gamma}
     \frac{dC_q}{d\tau}
     &= -\frac{C_q}{a}\int_{-\frac{\pi}{a}}^{\frac{\pi}{a}}dk k^2C_k-\frac{C_q q^2}{a}\int_{-\frac{\pi}{a}}^{\frac{\pi}{a}}dk C_k + \frac{1}{2\pi}\int_{-\frac{\pi}{a}}^{\frac{\pi}{a}}dkdp \ C_p C_k k^2 \nonumber \\
     &= (n-C_q/a)\int_{-\frac{\pi}{a}}^{\frac{\pi}{a}} dk k^2C_k - 2\pi C_qq^2 n/a.
\end{align}
Albeit all the remaining integrals in Eqs.~\eqref{eq:gamma_rate-sum} and \eqref{eq:rate-gamma} can be extended to the whole real line, since $C_q$ decays rapidly enough at large $q$, the resulting equation still explicitly depends on the lattice spacing $a$. This is a crucial difference compared to the case of $K$ body annihilation discussed before and it impacts on the amplitude of the density decay, as shown in Fig.~\ref{fig:conv}

In Fig.~\ref{fig:conv}(a), in particular, we plot the density $n(\tau)$ as a function of $\tau$ for various values of the lattice spacing $a$. The amplitude of the decay depends on the chosen value of $a$. The decay exponent, on the other hand, does not and the corresponding decay law is
\begin{equation}
n(\tau) \sim \tau^{-1}.
\label{eq:coagulation_continuum_exponent}
\end{equation}
This is shown in Fig.~\ref{fig:conv}(b), where the effective exponent $\delta(\tau)$ \eqref{def:delta} is plotted for various values of $a$. In all the cases $\delta(\tau)$ converges rapidly to $-1$. This result shows that in the reaction-limited regime binary annihilation and coagulation do not belong to the same universality class since they show asymptotic decays with different exponents \eqref{n_ann} and \eqref{eq:coagulation_continuum_exponent}, respectively. This in contrast with classical RD systems, where binary annihilation and coagulation are shown to display algebraic decay with analogous exponents \cite{henkel1995equivalences,henkel1997reaction,krebs1995finite}. For lattice fermionic systems the inequivalence between binary annihilation and coagulation had been observed in Ref.~\cite{perfetto22}. Here, we show that such inequivalence is not a lattice effect as it is robust with respect to taking the continuum limit. This surprising result is a direct consequence of the fermionic statistics and can be intuitively understood by looking at the structure of the jump operators \eqref{eq:binary_ann_continuum_FB} and \eqref{eq:coagulation_continuum}. The binary annihilation jump operator \eqref{eq:binary_ann_continuum_FB}, indeed, contains a pair of fermionic annihilation operators and it therefore accounts for the fermionic hard-core repulsion between the pair of nearby annihilated fermions only. The coagulation jump operator \eqref{eq:coagulation_continuum}, on the contrary, contains three fermionic operators. As a consequence, this operator feels the hard-core repulsion between the removed fermion and all the other fermions on its left (this simply follows by applying the definition for the action of a fermionic field operator $\psi(x)$ over a many-body state expressed in second quantization). The fermionic hard-core constraint in coagulation thereby induces an effective long-range repulsion which averages fluctuations over a larger number of repulsively interacting fermionic pairs. This intuitively explains the emergence of mean-field decay \eqref{eq:coagulation_continuum_exponent} for fermionic reaction-limited coagulation. We also note that in continuum space while for binary annihilation the amplitude of the decay \eqref{eq:n_fd} is universal, i.e., independent on the lattice spacing, for coagulation it is not as it depends on the UV cutoff. 
\begin{figure}[h]
    \centering
    \includegraphics[width=0.495\columnwidth]{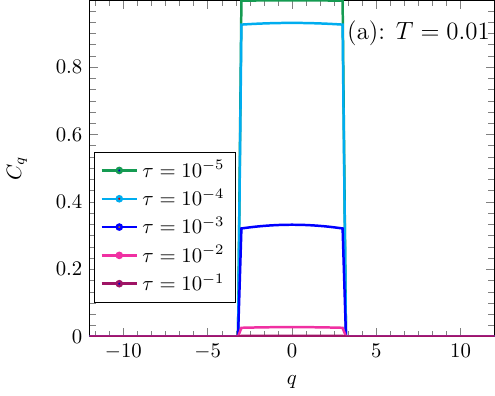}
     \includegraphics[width=0.495\columnwidth]{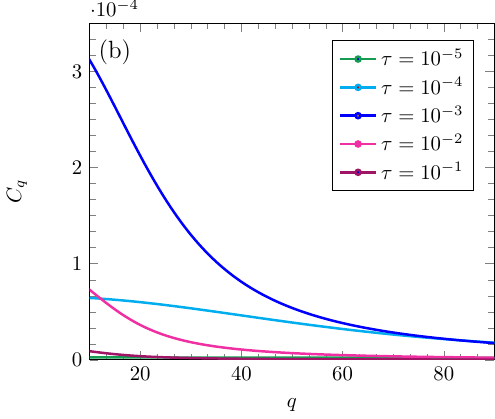}
    \caption{\textbf{Momentum occupation function for various values of time $\tau$ for coagulation.} (a) $C_q$ is plotted versus $q$ for values of the rescaled time $\tau$ (from top to bottom). The temperature of the initial state is $T=0.01$ and initial density $n_0=1$. The parameters used in the numerical solution of Eq.~\eqref{eq:rate-gamma} are $dk=0.1$ and $a=10^{-3}$. (b) Plot of high momentum values $q>10$. These momenta are initially not populated ($\tau=10^{-5}$ in the plot). At later times $\tau=10^{-4}, 10^{-3}$ those modes acquire a non-vanishing population, while they eventually decay to zero for later times ($\tau=10^{-2},10^{-1}$).} 
    \label{fig:c}
\end{figure}

In Fig.~\ref{fig:c}, we plot the occupation function $C_q(\tau)$ as a function of $q$ for various times and a fixed temperature $T=0.01$ (the same value used in Fig.~\ref{fig:conv}). In Fig.~\ref{fig:c}(a), one sees that similarly to the case of binary annihilation, $C_q$ becomes peaked around $q=0$, while higher momenta decay rapidly. Differently from annihilation, as shown in Fig.~\ref{fig:c}(b), initially unpopulated modes $q^{\star}$, such that $C_{q^{\star}}(0)=0$, can get populated as time progresses. This can be understood by inspecting the right hand side of \eqref{eq:gamma_rate-sum} which is not positive definite, differently from \eqref{eq:rate_alpha}. We also observe that the occupation of initially unpopulated modes is not monotonic in time, as it first increases to some non-zero value before eventually decaying to zero, as it must be since coagulation necessarily drives the system to a stationary state void of particles.

\begin{figure}[t]
    \centering
\includegraphics[width=0.495\columnwidth]{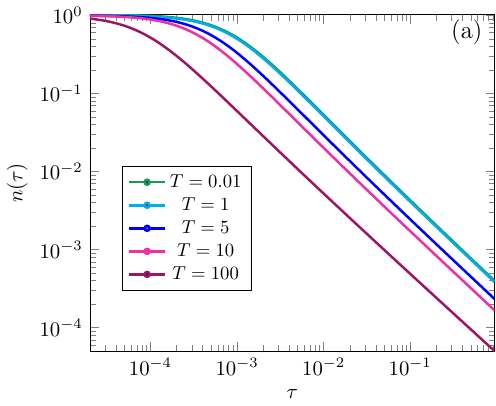}
\includegraphics[width=0.495\columnwidth]{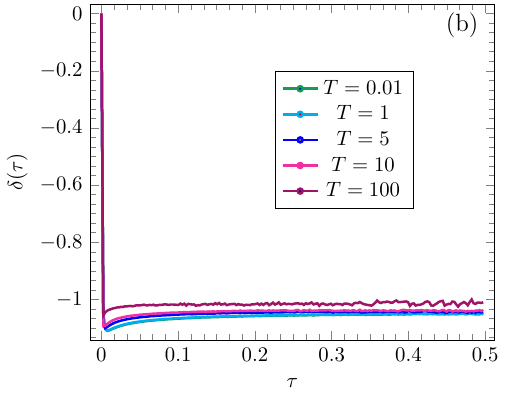}
    \caption{\textbf{Reaction-limited coagulation dynamics in the continuum for various values of the temperature.} (a) Log-log plot of the particle density $n(\tau)$ as a function of the rescaled time $\tau$ for increasing values of the temperature (from top to bottom). (b) Plot of the effective exponent $\delta(\tau)$ vs $\tau$. The parameters employed in the numerical solution of Eq.~\eqref{eq:rate-gamma} are $dk=0.1$ for the momentum grid and $a=2.5\cdot 10^{-3}$ for the UV cutoff $k_M=\pi/a$. Increasing the temperature $T$ accelerates the dynamics but it does not change the decay exponent $\delta\approx-1$. The different value to whom $\delta(\tau)$ converges at $T=100$ slightly deviates from that one attained by the other curves ($T=0.01,1,5,10$) due to numerical errors in the solution of Eq.~\eqref{eq:rate-gamma}. As a matter of fact, for large $T$ a larger cutoff $k_M=\pi/a$ and therefore finer grids (smaller $dk$) are needed since the occupation function is supported over a wider range of values in $q$. The initial density is $n_0=1$ for both panels.}
    \label{fig:gamma}
\end{figure}

In Fig.~\ref{fig:gamma}, we eventually plot the density $n(\tau)$ for various values of the temperature. Also in this case, upon increasing the temperature the amplitude of decay decreases rendering the decay faster. This is again caused by the fact that in continuum space, upon increasing the temperature higher momenta $q$ are excited, which decay rapidly. The decay exponent is, however, not affected as shown in Fig.~\ref{fig:gamma}(c). The decay law \eqref{eq:coagulation_continuum_exponent} is therefore a robust feature unaffected by statistical fluctuations in the initial state due to the temperature. We also checked that, as in the case of annihilation on a lattice, shown in Fig.~\ref{fig:lattice_binary_ann}, the decay approaches the mean-field prediction (both the amplitude and the exponent) as the temperature is increased. In the continuum, instead, the amplitude is never that of mean field and it depends on the lattice spacing.

\subsection{\textbf{Contact process and absorbing-state phase transition}}
\label{subsec:branching}
We study here the competition among branching, decay and binary annihilation. This leads to an absorbing-state phase transition, which we study in this Subsection. The TGGE rate equation for branching is obtained in a similar way as for the case of coagulation of Subsec.~\ref{subsec:coagulation}. We therefore do not report the intermediate steps for the sake of brevity, while the final expression reads as 
\begin{align}
     \frac{dC_q}{dt} &=\frac{a^4\Gamma_\beta}{L^2} \sum_{p,k }  [ C_kC_{p}(-p^2 + pk )  
+C_kC_q ( k^2- q^2 )
+ C_k(p^2+q^2)-C_qk^2].
\label{eq:branching_1}
\end{align}
In the continuum limit we again need to introduce a finite UV-cutoff $|k|<\pi/a$, as in Eq.~\eqref{eq:uv-cutoff}, in order to regularize the sums over the quasi-momentum $p$ in \eqref{eq:branching_1}, which would otherwise be divergent. With the definition of the regularization factor
\begin{align}
  M_a :=  \int_{-\frac{\pi}{a}}^{\frac{\pi}{a}} dk\ k^2 = \frac{2\pi^3}{3a^3},
\label{eq:M_regularized}
\end{align}
we then have in the limit $L\to \infty$
\begin{equation}
\label{eq:rate-beta_0}
\frac{dC_q}{d t}= \widetilde{\Gamma}_{\beta}\left[(C_q- a n) \left(\int_{-\frac{\pi}{a}}^{\frac{\pi}{a}} dk\ C_kk^2 -M_a\right)
+  2\pi q^2 n ( 1-C_q) +\frac{a}{2\pi} \left(\int_{-\pi/a}^{\pi/a} dk \, k C_k \right)^2 \right]\!.
\end{equation}
It is here important to note that in contrast to coagulation, in Eqs.~\eqref{eq:gamma_rate-sum} and \eqref{eq:rate-gamma}, for branching the occupation function $C_q$ does not decay in general to zero for large $q$ values. This is clearly caused by the fact that branching creates particles and thereby in the active phase it drives the system towards a stationary state with $n^{\mathrm{stat}}\neq 0$. Consequently in Eq.~\eqref{eq:rate-beta_0} (and in all the subsequent equations of this Subsection) we cannot extend the upper and lower limits of the integral to $+\infty$ and $-\infty$, respectively. All the momentum integrals are therefore restricted to the range $|k|<\pi/a$. As in the case of the previous Subsections, we consider here thermal initial states \eqref{eq:n_fd} such that $C_k(0)=C_{-k}(0)$ and therefore $C_k(t)=C_{-k}(t)$ (cf. the discussion before Eqs.~\eqref{eq:rate-alpha-sym} and \eqref{eq:rate-gamma}). Consequently the last term of Eq.~\eqref{eq:rate-beta_0} vanishes and we have the TGGE rate equation
\begin{align}
\label{eq:rate-beta}
     \frac{dC_q}{d t}= \widetilde{\Gamma}_{\beta}\left[(C_q- a n) \left(\int_{-\frac{\pi}{a}}^{\frac{\pi}{a}} dk\ C_kk^2 -M_a\right)
+  2\pi q^2 n ( 1-C_q)\right]. 
\end{align}
In the previous equations, we defined the rescaled rate in the continuum limit $\widetilde{\Gamma}_{\beta}=a^3 \Gamma_\beta/(2\pi)$. We use here a different rescaling than that in Eq.~\eqref{eq:branching_continuum}. This choice, indeed, allows to have identical units [$\mbox{length}^3 \,\, \mbox{time}^{-1}$] for $\widetilde{\Gamma}_{\beta}$ and $\widetilde{\Gamma}_{2\alpha}$ in \eqref{eq:binary_ann_continuum_FB}. In this way the two quantities can be directly compared when binary annihilation and branching are simultaneously present (cf. Eq.~\eqref{eq:adimensional_rates_CP} here below). As a matter of fact, we here allow for the competition between branching, two-body annihilation and single-particle decay \eqref{eq:one_body_decay_continuum}. Including the contributions from binary annihilation (derived in Subsec.\ref{subsec:2bresults}) and single body decay to the rate equation \eqref{eq:rate-beta}, one eventually obtains:
\begin{align}
\label{eq:rate-cp}
   \frac{d C_q}{d \tau}&= (C_q- a n) \left(\int_{-\frac{\pi}{a}}^{\frac{\pi}{a}} dk C_kk^2 -M_a\right)
+  2\pi q^2 n ( 1-C_q) - \frac{\Gamma_\delta}{\tilde\Gamma_\beta} C_q  \nonumber \\
&- \frac{\tilde\Gamma_{2\alpha}}{\tilde\Gamma_\beta}\int_{-\frac{\pi}{a}}^\frac{\pi}{a} dk (q-k)^2C_k(\tau)C_q(\tau),
\end{align}
with the definition of the rescaled time $\tau= \tilde\Gamma_\beta t$.
Integration over the momenta $q$ yields the equation for the density of particles
\begin{align}
\label{eq:rate-n-cp}
    \frac{d n}{d \tau} &=  n\left(  M_a -\int_{-\frac{\pi}{a}}^{\frac{\pi}{a}}dq \ q^2C_q  \right)-\frac{\Gamma_\delta}{\widetilde{\Gamma}_\beta} \, n -2\, \frac{\widetilde\Gamma_{2\alpha}}{\widetilde{\Gamma}_{\beta}} \, n \int_{-\frac{\pi}{a}}^\frac{\pi}{a} dk C_k k^2 \nonumber \\
    &= n \int_{-\frac{\pi}{a}}^{\frac{\pi}{a}}dq\ q^2 \left[ 1-\lambda -  C_q(1+\varepsilon) \right],   
\end{align}
where we defined the adimensional parameters 
\begin{align}
\lambda:= \frac{\Gamma_\delta}{\tilde\Gamma_\beta M_a} \quad \text{and} \quad \varepsilon:= 2\frac{\tilde\Gamma_{2\alpha}}{\tilde\Gamma_\beta}.
\label{eq:adimensional_rates_CP}
\end{align}
Here, $\lambda$ quantifies the relative strength between one-body decay $\Gamma_{\delta}$ and branching, where the regularization factor $M_a$ \eqref{eq:M_regularized} ensures that $\widetilde{\Gamma}_{\beta} M_a$ has the dimensions of an inverse time. The factor $\varepsilon$, similarly quantifies the relative strength between binary annihilation and branching. As the parameters $\lambda$ and $\varepsilon$ are tuned, one therefore obtains an absorbing-state phase transition. Namely since $C_q\geq0$, the right-hand side of Eq.~\eqref{eq:rate-n-cp}
can be positive, only if $\lambda<1$. Otherwise we have $\dot n<0$ and thus evolution towards the absorbing state.
This indicates an absorbing state phase transition with a critical point at 
\begin{equation} 
\lambda_c = \frac{\Gamma_\delta}{M_a\widetilde{\Gamma}_\beta} = \frac{3}{\pi^2} \frac{\Gamma_\delta}{\Gamma_\beta}=1
\label{eq:critical_point_CP}
\end{equation}
that is independent on binary annihilation strength $\widetilde{\Gamma}_{2\alpha}$. The critical ratio $\Gamma_{\delta}/\widetilde{\Gamma}_{\beta}$ depends on the UV cutoff, which shows that the critical point is a non-universal feature. For $\lambda\geq \lambda_c$, the only stationary state is the absorbing one with $n^{\mathrm{stat}}=0$ and $C_q^{\mathrm{stat}}=0$. For $\lambda<\lambda_c$, instead, the active phase is established and one obtains a nonvanishing stationary density $n^{\mathrm{stat}}\neq 0$ and occupation function $C_q^{\mathrm{stat}}$. The latter can be written in terms of the former by inserting  Eq.~\eqref{eq:rate-n-cp} into the rate equation ~\eqref{eq:rate-cp} and setting the time derivatives to zero
\begin{align}
\label{eq:c-stat}
C_q^\mathrm{stat}=n^\mathrm{stat} \frac{aM_a\left(1-\frac{1-\lambda}{1+\varepsilon} \right)+2\pi q^2}{ M_a\left( 1+\lambda - \frac{1-\lambda}{1+\varepsilon}(1+\varepsilon/2) \right)+2\pi q^2n^\mathrm{stat}\left(1+\varepsilon/2  \right)}.
\end{align}
In the absence of binary annihilation ($\varepsilon=0$) this simplifies to 
\begin{align}
\label{eq:c-stat-e=0}
C_q^\mathrm{stat}&=n^\mathrm{stat} \frac{a\lambda M_a+2\pi q^2}{2\lambda M_a+2\pi q^2n^\mathrm{stat}}.
\end{align}
\begin{figure}
    \centering
\includegraphics[width=0.495\columnwidth]{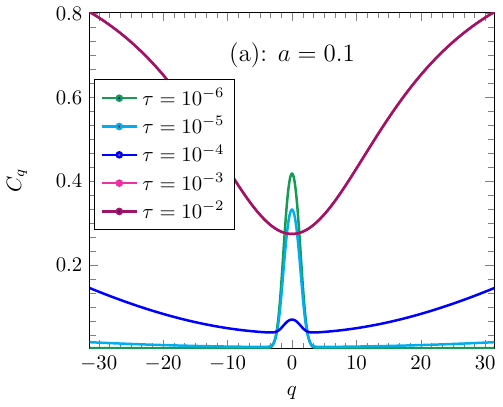}
\includegraphics[width=0.495\columnwidth]{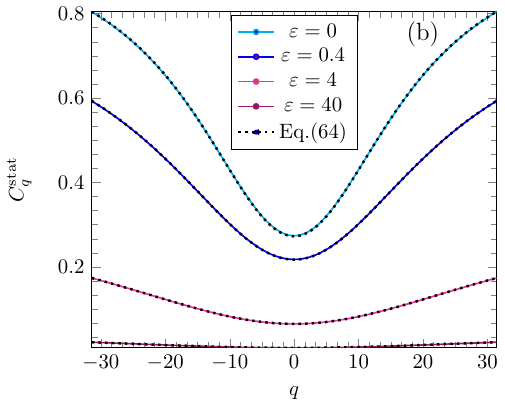}
    \caption{\textbf{Stationary momentum occupation function in the active phase of the contact process.} (a) Plot of $C_q$ (numerical solution of Eq.~\eqref{eq:rate-cp}) vs $q$ for various values of the rescaled time $\tau=\tilde\Gamma_\beta t$ for $a=0.1$, with UV cutoff $k_M=\pi/a$. At long times, $\tau=10^{-2}$, $C_q(\tau)$ converges towards the stationary state $C_q^\mathrm{stat}$ in the active phase. The latter matches the prediction in Eq.~\eqref{eq:c-stat-e=0}. The parameters are $\lambda=0.3$, $T=1$, $n_0=0.2$ in Eq.~\eqref{eq:n_fd}, and $\varepsilon=0$. The curves at $\tau=10^{-3}$ and $\tau=10^{-2}$ coincide. (b) Plot of $C_q^{\mathrm{stat}}$ as a function of $q$ for increasing values of $\varepsilon$ (from top to bottom). The other parameters are the same as in panel (a). The density $n^{\mathrm{stat}}$ that is needed for the analytic curves Eq.~\eqref{eq:c-stat} is obtained from the numerical solution for $C_q^{\mathrm{stat}}$.}
    \label{fig:stat-c_Q}
\end{figure}
We compare formula \eqref{eq:c-stat} with the numerical solution of Eq.~\eqref{eq:rate-n-cp} in Fig.~\ref{fig:stat-c_Q}. In Fig.~\ref{fig:stat-c_Q}(a) we plot $C_q(\tau)$ to study how the stationary distribution $C_q^{\mathrm{stat}}$ is approached as $\tau$ increases. We set $\varepsilon=0$ so that binary annihilation is absent. The result depends on the UV cutoff and $a$, which is necessary to ensure convergence of the integrals in \eqref{eq:rate-cp} for large $q$. Interestingly, we find that in the stationary state the mode $q=0$ has the minimal population, despite it being initially the one with the largest occupation according to Eq.~\eqref{eq:n_fd}. At long times ($\tau=10^{-2}$ in the figure), the stationary distribution $C_q^{\mathrm{stat}}$ matches with excellent agreement \eqref{eq:c-stat-e=0}. In Fig.~\ref{fig:stat-c_Q}(b), we plot $C_q^{\mathrm{stat}}$ for various values of $\varepsilon$. In particular, we observe that adding binary annihilation does not change the qualitative shape of $C_q^{\mathrm{stat}}$, which still features a minimal occupation around $q=0$, but only its width and amplitude. The latter decreases as $\varepsilon$ increases, as it must be since binary annihilation increasingly depletes the system.

It is interesting to compare the result in Fig.~\ref{fig:stat-c_Q} for the fermionic gas with the stationary state obtained in the lattice model in Ref.~\cite{perfetto22}. In the latter, the active stationary state, was found to feature spatial correlations, i.e., a dependence of $C_q^{\mathrm{stat}}$ on $q$, only when the binary annihilation process \eqref{eq:annihilation_interference} ($\theta\neq0,\pi$) is considered. Furthermore, on the lattice, the inverse Fourier transform of $C_q^{\mathrm{stat}}$ shows strong correlations at distance $x=2a$, which can be interpreted in terms of the dark states of the annihilation \eqref{eq:annihilation_interference} jump operator. In the continuum, the nature of the correlations in the stationary state is different. With and without ($\varepsilon=0$) annihilation, the inverse Fourier transform of $C_q^{\mathrm{stat}}$ reveals correlations at all distances $x$. The lattice dark state contribution to correlations at distance $x=2a$ is therefore a lattice effect which disappears in the continuum limit. For a fermionic quantum gas in continuum space, the qualitative structure of the stationary state $C_q^{\mathrm{stat}}$ is then dictated only by decay $A\to \emptyset$ and branching $A\to A+A$. This result is in agreement with classical RD systems \cite{vladimir1997nonequilibrium,hinrichsen2000non,henkel2008non,tauber2002dynamic,tauber2005,tauber2014critical}, where binary annihilation is needed to produce a finite stationary density in the active phase induced by branching. For fermions, however, the double occupancy restriction already necessarily makes the stationary density finite. This explains why the qualitative structure of the stationary state is not modified by the presence of additional binary annihilation. We now show that also the universality class of the stationary-state phase transition is not affected by the presence of binary annihilation.
\begin{figure}[h]
    \centering
\includegraphics[width=0.5\columnwidth]{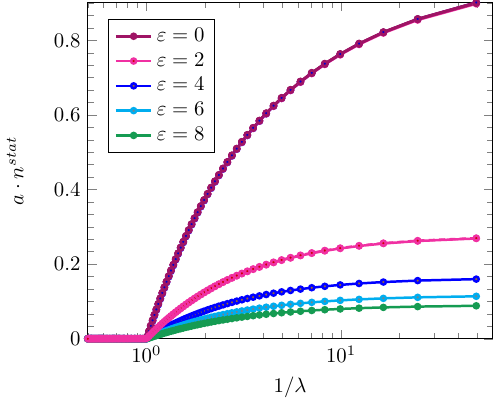}
    \caption{\textbf{Reaction-limited phase diagram of the fermionic contact process in the continuum space limit.} Plot of the stationary density $n^{\mathrm{stat}}$ as a function of $1/\lambda$ for increasing values of $\varepsilon$ (from top to bottom). The data are obtained from the numerical solution of Eq.~\eqref{eq:rate-cp}. The initial state is thermal at temperature $T=1$ and initial density $n_0=0.2$. The value $a=0.04$ is taken, which fixes the UV cutoff $k_M=\pi/a$. The stationary density on the vertical axis is multiplied by $a$ so that $a n^{\mathrm{stat}}$ is adimensional. The critical value $\lambda_c=1$ separates the absorbing phase ($\lambda>\lambda_c$ and $n^{\mathrm{stat}}=0$) from the active phase ($\lambda<\lambda_c$ and $n^{\mathrm{stat}}\neq 0$).}
    \label{fig:stat-c}
\end{figure}

From the stationary state occupation function we can eventually compute the stationary state density $n^{\mathrm{stat}}$. The corresponding data are shown in Fig.~\ref{fig:stat-c}. For $\lambda\geq \lambda_c$, the absorbing phase with $n^{\mathrm{stat}}=0$ is stabilized, while for $\lambda<\lambda_c$ the active phase $n^{\mathrm{stat}}$ is established. In agreement with Eq.~\eqref{eq:critical_point_CP} for the critical point $\lambda_c$, one sees that the latter does not depend on the binary annihilation strength $\varepsilon$. Binary annihilation only affects the value of the order parameter $n^{\mathrm{stat}}$ in the active phase, which gets lower as $\varepsilon$ is increased.  
At the critical point $\lambda_c$, the order parameter $n^{\mathrm{stat}}$ vanishes continuously and the transition is therefore of second order. The universality class of the transition can accordingly be classified by studying the scaling behavior of $n^{\mathrm{stat}}$ in the vicinity $\lambda\to \lambda_c$ of the critical point. We restrict ourselves to the case in Eq.~\eqref{eq:c-stat-e=0} with $\varepsilon=0$ for the sake of brevity. Integrating Eq.~\eqref{eq:c-stat-e=0} over $q$ gives 
\begin{align}
    n^\mathrm{stat}&=\frac{1}{2\pi}\int_{-\frac{\pi}{a}}^\frac{\pi}{a} C_q^\mathrm{stat}
    = \frac{1}{a} -\sqrt{\frac{\lambda M_a}{\pi^3n^\mathrm{stat}}}(1-an^\mathrm{stat}/2)\arctan\left[\sqrt{\frac{\pi^3n^\mathrm{stat}}{\lambda M_a}}\frac{1}{a} \right].
\end{align}
The latter is a trascendental equation for $n^{\mathrm{stat}}$, which cannot be exactly solved. We can, however, expand the right hand side close to the critical point, i.e., for $n\ll 1$: 
\begin{align}
   \sqrt{\frac{\pi^3n^\mathrm{stat}}{\lambda M_a}}\left(n^\mathrm{stat}-\frac{1}{a}\right) & \approx -(1-an^\mathrm{stat}/2)\left(\sqrt{\frac{\pi^3n^\mathrm{stat}}{\lambda M_a}}\frac{1}{a}-\frac{1}{3}\left(\frac{\pi^3n^\mathrm{stat}}{\lambda M_a}\frac{1}{a^2}\right)^{\frac{3}{2}} \right).
\end{align}
From this result we can immediately read the asymptotic behavior of $n^{\mathrm{stat}}$ as $\lambda\to \lambda_c$
\begin{align}
   n^\mathrm{stat} = \left\lbrace\begin{array}{cc}
    \frac{2}{a}\left( 1-\frac{3\pi^3\lambda}{2a^3M_a}\right)\sim (\lambda_c-\lambda)^1,  &  \lambda <\lambda_c \\
     0, & \lambda \geq \lambda_c
\end{array}\right..
\label{eq:beta-exp}
\end{align}
Thus we find that the stationary density $n^{\mathrm{stat}}(\lambda)$ follows a power law in $\lambda$ as $\lambda \to \lambda_c$. The associated exponent is $\beta=1$ and it does not depend on the UV cutoff. A comparison with the stationary density obtained from the numerical solution of \eqref{eq:rate-cp} shows excellent agreement. From the numerical solution of Eq.~\eqref{eq:rate-cp} we can also determine the critical exponent associated with the decay of the density when $\lambda=\lambda_c$. In this case, we find the power law in time
\begin{equation}
n(\tau) \sim \tau^{-1},
\label{eq:delta_exp_CP}
\end{equation}
which is also independent on the UV cutoff. Both the exponents \eqref{eq:beta-exp} and \eqref{eq:delta_exp_CP} coincide with the exponents of classical directed percolation within the mean-field approximation \cite{hinrichsen2000non,vladimir1997nonequilibrium,henkel2008non,tauber2014critical,Malte_Henkel_2004}. The same universality class has been found for lattice fermionic model in Ref.~\cite{perfetto22} and we conclude that the universality class of the absorbing-state phase transition of the Fermi gas is not changed by the continuum limit. In the reaction-limited regime, the critical exponents associated to the phase transition are those of mean-field directed percolation. The mean-field values for the critical exponents \eqref{eq:beta-exp} and \eqref{eq:delta_exp_CP} in one spatial dimension might look surprising. For the classical contact process, one, indeed, finds that the critical exponents associated to the stationary-state phase transition deviate from the mean-field values for spatial dimensions $d$ smaller than the so-called upper critical dimension $d_c=4$ \cite{henkel2008non,tauber2005,tauber2014critical}. It is, however, fundamental to highlight that the upper critical dimension $d_c$ characterizes the \textit{diffusion-limited} regime (strong dissipation, not analyzed in this work). In the latter, mixing of the particles is slow and spatial fluctuations in the density profile are relevant in low space dimensions. This leads to the concept of upper critical dimension $d_c$, below which critical exponents are different from their mean-field values. In the present work, we, instead, consider the \textit{reaction-limited} regime (weak dissipation). In this regime, the Hamiltonian mixing of the particles in-between consecutive reactions is strong and therefore spatial fluctuations are wiped out. Consequently, in the reaction-limited regime mean-field exponents can be observed already in one dimension.

\section{Conclusions}
\label{sec:V_conclusions}
In this manuscript, we studied the Fermi gas undergoing weak dissipative reactions in one spatial dimension. In particular, we focused on the comparison between the dynamics of the gas in continuum space and that of the lattice counterpart. The main finding of the manuscript is that long-time universal properties characterized by critical exponents coincide in the continuum gas and in the discrete lattice formulation. The latter only impact the nonuniversal properties of the dynamics such as the amplitude of the decay, the value of the critical point in the absorbing-state stationary phase transition and the associated stationary state. 

In our study we considered various reaction processes, such as binary \eqref{eq:binary_ann_continuum_FB}, three-body \eqref{eq:3ann_continuum} annihilation, coagulation \eqref{eq:coagulation_continuum} and branching \eqref{eq:branching_continuum}. We first derived the quantum master equation for the gas in continuum space in Eqs.~\eqref{eq:free_hamiltonian_continuum} and \eqref{eq:dissipator_continuum} by taking the limit of vanishing lattice spacing $a\to 0$ in the corresponding lattice master equation. In doing so, we highlighted how the continuum limit of coagulation and branching requires normal ordering since the corresponding jump operators are composed of creation and annihilation operators which do not anticommute in the continuum limit. Based on this formulation, we investigated the dynamics in the reaction limited regime of weak dissipation. To this end we adopted the time-dependent generalized Gibbs ensemble method \cite{TGGE1,TGGE2,TGGE3,TGGE4,TGGE5,TGGE6}, which leads to a large-scale kinetic equation \eqref{eq:tGGE-rate} for the occupation function $C_q$ in momentum space $q$. 

\begin{table}[h]
    \centering
    \begin{tabular}{|c|c|c|c}
      \hline
        RD Process  & Lattice & Continuum \\\hline
        $2\alpha$  & $n(\tau)\sim \tau^{-1/2}$ or $n(\tau)\sim \tau^{-1}$  & $n(\tau)\sim (T\tau)^{-1/2}$   \\  \hline
        $3\alpha$  &  non algebraic or $n(\tau)\sim \tau^{-1/2}$ & non algebraic  \\ \hline
        $\gamma$  &  \multicolumn{2}{|c|}{$n(\tau)\sim \tau^{-1}$}    \\ \hline
        $\delta$, $\beta$,  $2\alpha$ &  \multicolumn{2}{|c|}{ $n^{\mathrm{stat}}\sim (\lambda_c-\lambda)^1$, and $n(\tau) \sim \tau^{-1}$}   \\ \hline
    \end{tabular}
    \caption{\textbf{Summary of quantum reaction-limited scaling behavior for the Fermi gas. Lattice versus continuum comparison}. On each row the considered RD model is indexed by the index $\nu=2\alpha, 3\alpha,\gamma,\delta,\beta$ according to the notation of Sec.~\ref{sec:II_continuum}. For binary and three-body annihilation the main difference between lattice and continuum lies in the influence of the temperature $T$ on the asymptotic decay. On the lattice, when the temperature is increased, one finds mean-field decay $n(\tau)\sim \tau^{-1}$ and $n(\tau)\sim \tau^{-1/2}$ for binary and three-body annihilation, respectively. For coagulation, mean-field decay is found both on the lattice and in the continuum regardless of the temperature. For the contact process model with binary annihilation, critical exponents are the same on the lattice and on the continuum. The value of the critical coupling $\lambda_c$ is nonuniversal and it is given in Eq.~\eqref{eq:critical_point_CP} in terms of the ratio $\widetilde{\Gamma}_{\beta}/\Gamma_{\delta}$ for the continuum model, while it is equal to $\lambda_c=\Gamma_{\beta}/\Gamma_{\delta}=1$ for the lattice.}
    \label{tab:exponents}
\end{table}

We solved this equation for the various dissipative processes considered. Our results are summarized in Table \ref{tab:exponents}. In the case of binary annihilation $2A\to \emptyset$ we extended previous results in Refs.~\cite{Rosso2022,gerbino2023largescale,gerbino2024kinetics,maki2024loss}, valid for zero-temperature initial states, to generic values of the temperature $T$ of the initial state. This parameter does not impact on the asymptotic decay exponent $n(\tau)\sim (\tau T)^{-1/2}$, but only on the amplitude of the decay, as shown in Eq.~\eqref{eq:n_th_a}. This result is qualitatively different from the corresponding lattice dynamics, where mean-field decay $n(\tau)\sim \tau^{-1}$ is found in an initial time window which extends as the temperature is increased (see Fig.~\ref{fig:lattice_binary_ann}(a)). We explained this important difference between the lattice and continuum case by noting that in the latter high temperatures lead to faster particles and therefore faster mixing and decay. In the former, instead, the maximal propagation velocity is fixed, then high temperatures lead to maximal entropy state with homogeneous filling of the available velocities. Such states lack spatial correlations and therefore they lead to mean-field decay in the reaction-limited regime. We then discussed the case of three-body annihilation $3A\to\emptyset$ in Eq.~\eqref{eq:3ann_kin_eq}. The fermionic lattice model had been previously discussed in Ref.~\cite{perfetto23}, while here we first discuss the continuum limit. In the lattice formulation of $3A\to \emptyset$, the temperature $T$ of the initial state has a similar influence as in the case of binary annihilation. Also for $3A\to\emptyset$, indeed, one finds mean-field decay $n(\tau)\sim \tau^{-1/2}$ in an initial time interval which increases as the initial-state temperature is raised. The crucial difference compared to $2A\to\emptyset$ is that at longer times, instead, a non algebraic decay was found in \cite{perfetto23}. In the continuum limit, we also find a non algebraic decay at long times regardless of the initial temperature $T$. In the case of coagulation $2A\to A$, we here first derive the corresponding TGGE rate equation \eqref{eq:gamma_rate-sum-0}. This equation is regularized by inserting an UV cutoff in order to avoid divergences at high momenta in the continuum limit. From the regularized equation \eqref{eq:rate-gamma}, we found that the long-time decay follows the mean-field exponent $n(\tau) \sim \tau^{-1}$, as well as in the lattice case. We eventually considered the absorbing-state phase transition induced by the competition between branching $A\to 2A$, one-body decay $A\to \emptyset$ and $2A\to \emptyset$. This model had formerly been discussed in Ref.~\cite{perfetto22} on the lattice. Here we address the effect of the continuum limit on the associated stationary state and scaling behavior. For the branching jump operator, as in the case of coagulation, one needs to include a UV cutoff. From \eqref{eq:rate-cp} we found that the associated critical exponents in Eq.~\eqref{eq:delta_exp_CP} belong to the mean-field directed percolation universality class. The continuum limit impacts, however, on the stationary state and the associated correlations. The latter are present even for zero annihilation rate, differently from the lattice case, and they occur at all distances.       

Beyond this analysis, it is further interesting to look at the case of spinful fermions, as in the Fermi-Hubbard model \cite{lossexpF2,lossexpF3,FHth1,FHth2,FHth3}, both in the lattice and in the continuum. Models of spinful fermions with dissipation naturally link to reaction-diffusion processes, such as $A+B\to \emptyset$ binary annihilation of the two species upon meeting. The ensuing dynamics for classical RD models has been shown to lead to a rich decay depending on whether the species $A$ and $B$ initially have equal or unequal densities, and whether they are initially homogeneously distributed or not \cite{toussaint1983particle,kang84scaling,meakin1984novel,kang85,BramsonSegregated,howard1996,tauber2005}. In the quantum case, we similarly expect a rich dynamics, which can be addressed by including spatial inhomogeneities in the initial state into the present analysis. In a related way, it is also interesting to study weakly dissipative spin $1/2$ lattice systems. Recently, in Ref.~\cite{riggio2023}, these systems have been analyzed via the same TGGE method of the manuscript. In the case of even dissipative processes, which does not change the parity of particle number, such as $2A\to \emptyset$, spin $1/2$ systems in $d=1$ can be simply mapped onto fermions via Jordan-Wigner transformation. For odd processes, such $3A\to \emptyset$, $2A\to A$ and $A\to 2A$ considered here, however, nonlocal fermionic strings are introduced by the Jordan-Wigner transformation. It is then natural to wonder whether such terms alter the critical exponents of the dynamics both in the case of coagulation and for the stationary-state phase transitions of the contact process. We plan to address these questions in future studies.

\ack{We acknowledge funding from the Deutsche Forschungsgemeinschaft within the Grant No. 449905436 and the research units FOR5413 (Grant No. 465199066) and FOR5522 (Grant No. 499180199). We also acknowledge funding from the Bundesministerium f\"ur Forschung und Bildung (F\"orderprogramm: Quantensysteme, F\"orderkennzeichen: 13N17065). Funding was also received from EPSRC (Grant No. EP/V031201/1).}

\appendix
\section{Binary annihilation dynamics}
\label{app:Appendix1}
We provide here additional details about the continuum limit of the jump operator \eqref{eq:annihilation_interference}. In addition, we also provide the intermediate steps leading to the asymptotics at long times for lattice binary annihilation at high, Eq.~\eqref{eq:l_as_th}, and low, Eq.~\eqref{eq:l_as_0}, temperatures. 

The continuum limit of Eq.~\eqref{eq:annihilation_interference} leads to \eqref{eq:binary_ann_continuum_FB} for $\theta \neq \theta^{\star}$, with $\theta^{\star}= -\pi/4 +\pi \mathbb{Z}$. In the latter case, additional care must be taken since the rescaled rate $\widetilde{\Gamma}_{2\alpha}$ in Eq.~\eqref{eq:binary_ann_continuum_FB} vanishes. We therefore need to consider the next order in the expansion in $a$:
\begin{align}
    L^\alpha(x) &=\lim_{a\to 0} \frac{1}{\sqrt{a}} L^{2\alpha}_{ja}(\theta^{\star})= \frac{1}{\sqrt{2a}}\sqrt{a^2\Gamma_{2\alpha}}\psi(x)\left[ \left(\psi (x) +a\partial_{x}\psi(x) + \frac{a^2}{2}\partial_{x}^2\psi(x) \right) \right. \nonumber \\
    &\qquad \qquad \qquad \qquad \qquad \qquad \quad +\left. \left(\psi (x) -a\partial_{x}\psi(x)+ \frac{a^2}{2}\partial_{x}^2\psi(x) \right)\right] \nonumber \\
     &=\sqrt{\widetilde{\Gamma}_{2\alpha}^{\theta^{\star}}}\psi(x) \partial_{x}^2\psi(x), \quad \mbox{with} \quad \widetilde{\Gamma}_{2\alpha}^{\theta^{\star}}=\frac{a^5\Gamma_{2\alpha}}{2},
\label{supeq:continuum_theta_star}
\end{align}
where we used that $|\cos \theta^{\star}-\sin \theta^{\star}|=\sqrt{2}$. From Eq.~\eqref{supeq:continuum_theta_star} one can then obtain the associated TGGE rate equation \eqref{eq:tGGE-rate} following similar steps as in Eq.~\eqref{eq:rate_alpha_intermediate}, which reads as
\begin{equation}
\label{eq:rate_alpha_theta}
    \frac{dC_q(\tau)}{d\tau'} = -\int_{-\infty}^\infty dk (q^2-k^2)^2C_k(\tau')C_q(\tau'),
\end{equation}
with the rescaled time $\tau := t\dfrac{a^5\Gamma_\alpha}{\sqrt{2^3}\pi}$. We solved numerically Eq.~\eqref{eq:rate_alpha_intermediate} and we checked that the asymptotic decay of the density as a function of time is analogous to the case $\theta \neq \theta^{\star}$: $n(\tau) \sim \tau^{-1/2}$. In a similar way, we also checked that the temperature $T$ of the initial state does not change the asymptotic exponent, but only increases the amplitude of the decay similarly to what is shown in Fig.~\ref{fig:alpha}(a). The effective exponent $\delta(\tau)$ is also found to converge more rapidly to the asymptotic value $\delta=-1/2$ as the temperature $T$ is increased. The asymptotic decay of the Fermi gas in the presence of weak binary annihilation is therefore unaffected by the value chosen for $\theta$ in the original lattice formulation in Eq.~\eqref{eq:annihilation_interference}.

Regarding the lattice formulation for $\theta=\pi/4$, we also present here the steps missing in the derivation of the asymptotics \eqref{eq:l_as_th} and \eqref{eq:l_as_0}. We start by considering the chemical potential $\mu(\beta)$, that is determined by fixing the initial filling $n_0$
\begin{align}
\label{eq:filling}
n_0 &= \frac{1}{2\pi}\int_{-\pi}^\pi dq \frac{1}{1+e^{-\beta \cos q-\beta\mu}}.
\end{align}
In the limit of high temperature, i.e., $\beta\rightarrow 0$, the product $\beta\mu$ approaches for the lattice system a constant while $\beta\cos q\rightarrow 0$. We can then perform a Taylor expansion of the integrand in $x=\mbox{exp}(-\beta \cos q)$ around $x=1$ to leading order $x-1$. Integrating term by term the series as in Eq.~\eqref{eq:filling}, one has
\begin{align}
n_0 = \frac{1}{2\pi}\int_{-\pi}^\pi dq \frac{1}{1+e^{-\beta \cos q-\beta\mu}}= \frac{1}{\left(1+e^{-\beta\mu}\right)^2}\left( 1+2e^{-\beta\mu}- e^{-\beta\mu}I_0(\beta) \right).
\end{align}
Here, $I_0(\beta)$ denotes the modified Bessel function of first kind \cite{NIST:DLMF}. At leading order in $\beta \to 0$, the latter expands as $I_0(\beta\to0)\approx 1+\beta^2/4+\mathcal{O}(\beta^4)$ and we have \begin{align}
  n_0 &= \frac{1}{\left(1+e^{-\beta\mu}\right)^2}\left( 1+e^{-\beta\mu}- e^{-\beta\mu}\frac{\beta^2}{4} +\mathcal{O}(\beta^4) \right).
\end{align}
The previous equation is readily inverted in order to express $\mbox{exp}(-\beta \mu)$ as a function of $n_0$ (it amounts to solving a second-order algebraic equation in $\mbox{exp}(-\beta \mu)$) and it leads to 
\begin{align}
     e^{-\beta\mu} = \frac{1-n_0}{n_0}\left(1-\frac{\beta^2}{4}\right) +\mathcal{O}(\beta^4).
\label{supeeq:taylor_high_T}
\end{align}
We eventually insert \eqref{supeeq:taylor_high_T} in the expressions of $C_0$ and $C_{\pi}$, Eq.~\eqref{eq:asymp_lattice}, to get the leading dependence of the initial occupation function on the inverse temperature $\beta$
\begin{align}
    C_0 &= \frac{1}{1+e^{-\beta(\mu+1)}} = \frac{1}{1+\frac{(1-n_0)}{n_0}(1-\beta) +\mathcal{O}(\beta^2) } \approx \frac{n_0}{1-\beta(1-n_0)}, \\
    C_\pi &= \frac{1}{1+e^{-\beta(\mu-1)}} = \frac{1}{1+\frac{(1-n_0)}{n_0}(1+\beta) +\mathcal{O}(\beta^2) } \approx \frac{n_0}{1+\beta(1-n_0)}.
\end{align}
Inserting the latter equation into Eq.~\eqref{eq:asymp_lattice} one eventually obtains \eqref{eq:l_as_th}. The low-temperature asymptotics is, instead, immediate to obtain from Eq.~\eqref{eq:asymp_lattice} since as $\beta \to \infty$, one has that $C_0 \to 1$, while $C_{\pi}\to 0$. This leads to Eq.~\eqref{eq:l_as_0}. 
\section{Coagulation dynamics}
\label{app:Appendix3}
Here we report the calculations leading to Eq.~\eqref{eq:gamma_rate-sum-0} of the main text for the coagulation dynamics of the Fermi gas. We start by taking the Fourier transform \eqref{eq:fourier1} of the coagulation jump operator \eqref{eq:coagulation_continuum}
\begin{equation}
    L^{\gamma\pm}(x) = \sqrt{\frac{a^4\Gamma_\gamma}{L^3}}\sum_{p,p',p''}(-ip'')\phi^\dagger(p)\phi(p')\phi(p'')e^{ix(p-p'-p'')}.
\end{equation}
From the canonical anticommutation relations $\{\phi(k),\phi^{\dagger}(k') \}=\delta_{k,k'}$, we then compute the commutator
\begin{align}\label{eq:com-gamma}
    [\phi^\dagger(q)\phi(q), \phi^\dagger(p)\phi(p')\phi(p'')] = \phi^\dagger(p)\phi(p')\phi(p'')(\delta_{p,q}-\delta_{q,p'}-\delta_{q,p''}),
\end{align}
which we insert into the TGGE rate Eq.~\eqref{eq:tGGE-rate}. We accordingly get for the time derivative of $C_q$ that
\begin{align}
     \frac{dC_q}{dt} =&  \int_0^L dx \left\langle L^{\gamma\pm \dagger}(x) [ n(q), L^{\gamma \pm} (x) ] \right\rangle_\mathrm{GGE} (t)\nonumber \\
     =& \frac{\widetilde{\Gamma}_{\gamma}}{L^3}\sum_{p,p',p'',k,k'k'' } p''k''\left\langle \phi^\dagger(k'')\phi^\dagger(k')\phi(k) [ n(q), \phi^\dagger(p)\phi(p')\phi(p'') ] \right\rangle_\mathrm{GGE}\cdot \nonumber \\
     &\qquad \qquad \qquad \qquad \qquad \qquad \qquad \qquad \quad \quad \quad \quad \cdot \int_0^L dx e^{ix(p-p'-p''- k+k'+k'')}\nonumber\\
      = & \frac{\widetilde{\Gamma}_\gamma}{L^2} \sum_{p,p',p'',k,k'k'' } p''k''\left\langle \phi^\dagger(k'')\phi^\dagger(k')\phi(k) \phi^\dagger(p)\phi(p')\phi(p'')\right\rangle_\mathrm{GGE}\cdot \nonumber \\
      &\qquad \qquad \qquad \qquad \qquad \qquad \qquad \quad \quad  \cdot (\delta_{p,q}-\delta_{q,p'}-\delta_{q,p''}) \delta_{p-p'-p'', k-k'-k''}\nonumber \\
     =& \frac{\widetilde{\Gamma}_\gamma}{L^2} \sum_{p,p',p'',k,k'k'' } p''k''\left[ \left\langle \phi^\dagger(k'')\phi^\dagger(k')\phi(p')\phi(p'')\right\rangle_\mathrm{GGE}\delta_{k,p}\right.\nonumber \\  &- \left.\left\langle \phi^\dagger(k'')\phi^\dagger(k') \phi^\dagger(p)\phi(k)\phi(p')\phi(p'')\right\rangle_\mathrm{GGE} \right](\delta_{p,q}-\delta_{q,p'}-\delta_{q,p''}) \delta_{p-p'-p'', k-k'-k''}.
\label{supeq:intermediate_coagulation}
\end{align}
In passing from the second to the third equality we used Eq.~\eqref{eq:com-gamma} and the Fourier representation of the Kronecker delta function. In passing from the third to fourth equality we used the anticommutation relation between $\phi(k)$ and $\phi^{\dagger}(p)$. In this way we can express the resulting six-point fermionic correlation function, which is not normal ordered for coagulation, as the algebraic sum between a normal-ordered four-point function and a normal-ordered  six-point function. We then first evaluate the latter, which yields three-body terms with the product of three occupation functions in momentum space. Importantly we find that all such three-body terms exactly vanish:
\begin{align}
 \sum_{p,p',p'',k,k'k'' } p''k''& \left\langle \phi^\dagger(k'')\phi^\dagger(k') \phi^\dagger(p)\phi(k)\phi(p')\phi(p'')\right\rangle_\mathrm{GGE} \cdot \nonumber \\
 & \qquad \qquad \qquad \qquad \qquad \qquad \cdot (\delta_{p,q}-\delta_{q,p'}-\delta_{q,p''}) \delta_{p-p'-p'', k-k'-k''} \nonumber \\
 =\!\!\!\!\sum_{p,p',p'',k,k'k'' } \!\!\!\! p''k''&C_{k''}C_{k'}C_{p}[\delta_{k,p}(\delta_{k',p'}\delta_{k'',p''}-\delta_{k',p''}\delta_{k'',p'}) -\delta_{k',k}(\delta_{p,p'}\delta_{k'',p''}-\delta_{p,p''}\delta_{k'',p'})\nonumber \\
 &+\delta_{k,k''}(\delta_{p,p'}\delta_{k',p''}-\delta_{p,p''}\delta_{k',p'})] 
 (\delta_{p,q}-\delta_{q,p'}-\delta_{q,p''}) \delta_{p-p'-p'', k-k'-k''}\nonumber \\
 =\!\!\!\! \sum_{p,p',p'',k,k'k'' }\!\!\!\! p''k''&C_{k''}C_{k'}C_{p}[\delta_{k,p}(\delta_{k',p'}\delta_{k'',p''}-\delta_{k',p''}\delta_{k'',p'}) -\delta_{k',k}(\delta_{p,p'}\delta_{k'',p''}-\delta_{p,p''}\delta_{k'',p'}) \nonumber \\
 &\qquad \qquad +\delta_{k,k''}(\delta_{p,p'}\delta_{k',p''}-\delta_{p,p''}\delta_{k',p'})]\delta_{q,p} \delta_{p-p'-p'', k-k'-k''} \nonumber \\
 = \sum_{p'',k'k'' } p''k''&C_{k''}C_{k'}C_{q}[(\delta_{k'',p''}-\delta_{k',p''}) -(\delta_{k'',p''}-\delta_{q,p''})+(\delta_{k',p''}-\delta_{q,p''})]=0. 
\end{align}
Here we used Wick's theorem to decompose the fermionic six-point function in terms product of two-point functions only. In passing from the second to the third equality, we also renamed $k'\leftrightarrow p$. As a consequence, in Eq.~\eqref{supeq:intermediate_coagulation}, the only contribution is given by the fermionic normal ordered four-point function. The latter is computed in a similar way as in Eq.~\eqref{eq:rate_alpha_intermediate} and hence we obtain
\begin{align}
     \frac{dC_q}{dt} =& \frac{\widetilde{\Gamma}_\gamma}{L^2} \sum_{p,p',p'',k,k'k'' } p''k''\left\langle \phi^\dagger(k'')\phi^\dagger(k')\phi(p')\phi(p'')\right\rangle_\mathrm{GGE}\delta_{k,p} (\delta_{p,q}-\delta_{q,p'}-\delta_{q,p''})\cdot \nonumber \\
     & \qquad \qquad \qquad \qquad \qquad \qquad \qquad \qquad \qquad \qquad \qquad \qquad \cdot \delta_{p-p'-p'', k-k'-k''} \nonumber \\
     = & -\frac{\widetilde{\Gamma}_\gamma}{L^2}\sum_{k,p} (q-k)^2C_kC_q +  \frac{\widetilde{\Gamma}_\gamma}{L^2} \sum_{p,p',p'',k'k'' } p''k'' \left\langle \phi^\dagger(k'')\phi^\dagger(k')\phi(p')\phi(p'')\right\rangle_\mathrm{GGE} \delta_{p,q}\cdot  \nonumber \\
     & \qquad \qquad \qquad \qquad \qquad \qquad \qquad \qquad \qquad \qquad \qquad \qquad \qquad \cdot \delta_{p'+p'',k'+k''} \nonumber \\
     =& -\frac{\widetilde{\Gamma}_\gamma}{L^2}\sum_{k,p} (q-k)^2C_kC_q +  \frac{\widetilde{\Gamma}_\gamma}{L^2} \sum_{p',p'',k'k'' } p''k''C_{k'}C_{k''}(\delta_{k',p'}\delta_{k'',p''}-\delta_{k'',p'}\delta_{k',p''} ) \cdot \nonumber \\
     &\cdot \qquad \qquad \qquad \qquad \qquad \qquad \qquad \qquad \qquad \qquad \qquad \qquad \cdot \delta_{p'+p'',k'+k''} \nonumber \\
     =& -\frac{\widetilde{\Gamma}_\gamma}{L^2}\left[\sum_{k,p} (q-k)^2C_kC_q +\sum_{k,p} p(k-p)C_kC_p \right] \nonumber \\
     =& -\frac{\widetilde{\Gamma}_\gamma}{L^2}\left[\sum_{k,p} (q-k)^2C_kC_q +\frac{1}{2}\sum_{k,p} (k-p)^2C_kC_p \right].
\label{supeq:coagulation_final}
\end{align}
This equation coincides with Eq.~\eqref{eq:gamma_rate-sum-0} of the main text. A very similar derivation can be also carried on for the branching jump operator \eqref{eq:branching_continuum}. Also in this case we find that all the three-body terms involving products of three momentum occupation functions vanish. The remaining two-body terms after manipulations similar to those in Eq.~\eqref{supeq:coagulation_final} yield Eq.~\eqref{eq:branching_1} of the main text.

\section*{References}
\bibliography{refs}

\end{document}